\documentclass[journal]{IEEEtran}
\IEEEoverridecommandlockouts
\usepackage{cite}
\usepackage{multirow}
\usepackage{float}
\usepackage{stfloats}
\usepackage{enumitem}
\usepackage{amsmath,amsfonts}
\usepackage{amssymb}
\usepackage{bbm}
\usepackage{url}
\usepackage{nomencl}
\usepackage{algorithmic}
\usepackage{graphicx}
\usepackage{textcomp}
\usepackage[justification=centering]{caption}
\usepackage{xcolor}
\usepackage{booktabs}
\usepackage{etoolbox}
\usepackage{ntheorem}
\usepackage{makecell}
\usepackage{subfigure}
\theoremseparator{:}
\newtheorem{theorem}{Theorem}

\newtheorem{Corollary}{Corollary}

\def\BibTeX{{\rm B\kern-.05em{\sc i\kern-.025em b}\kern-.08em
    T\kern-.1667em\lower.7ex\hbox{E}\kern-.125emX}}
\begin{document}
\title{Integrating Renewable Energy Sources as Reserve Providers: Modeling, Pricing, and Properties}

%\author
\author{\hspace{-1.5em}Wenli~Wu,~\IEEEmembership{Student Member,~IEEE,}
        Ye Guo*,~\IEEEmembership{Senior Member,~IEEE,}%
       ~and~Jiantao Shi,~\IEEEmembership{Student Member,~IEEE}\vspace{-2em}%
        
\thanks{This work is supported in part by the National Science Foundation of China under Grant 52377105. Corresponding author: Ye Guo, e-mail: guo-ye@sz.tsinghua.edu.cn.}% 
}

\maketitle

\begin{abstract}
In pursuit of carbon neutrality, many countries have adopted renewable portfolio standards to facilitate the integration of renewable energy. However, increasing penetration of renewable energy resources will also pose higher requirements on system flexibility. Allowing renewable themselves to participate in the reserve market could be a viable solution. To this end, this paper proposes an optimal dispatch model for joint energy-reserve procurement that incorporates renewable portfolio standards and RES serve as reserve providers. Potential generator outages and deviations in renewable and load power are modelled through a given number of probability-weighted scenarios. In particular, reserve resources are initially booked in the base case and then activated in non-base scenarios through the re-dispatch process. Marginal pricing is used to derive energy, reserve, and power deviation prices. Next, we develop the associated settlement process and establish several market properties. The proposed pricing scheme establishes equivalence between thermal generators and renewable units by accounting for their uncertainties, including thermal generator outages and renewable power deviations, and their flexibility, namely reserve and re-dispatch. We have shown that for renewable resources, supplying reserve according to the dispatch results compared to generating as much as possible leads to better profits. Simulations validate the effectiveness of the proposed method and properties established.
\end{abstract}

\IEEEpeerreviewmaketitle
\vspace{-2em}\makenomenclature
\renewcommand\nomgroup[1]{%
  \item[\bfseries
  \ifstrequal{#1}{A}{Abbreviations}{%
  \ifstrequal{#1}{I}{Indices and Sets}{%
  \ifstrequal{#1}{P}{Parameters}{%
  \ifstrequal{#1}{V}{Variables}{%
  \ifstrequal{#1}{Q}{Variables}{}}}}}%
]}

\mbox{}
\nomenclature[A]{\(\text{RES}\)}{Renewable Energy Sources}
\nomenclature[A]{\(\text{RPS}\)}{Renewable Portfolio Standard}
\nomenclature[I]{\(i\)}{Index of generator}
\nomenclature[I]{\(b_{i}\)}{Bus index where generator $i$ is located}
\nomenclature[I]{\(k \in \mathcal{K}\)}{Index of scenarios}
\nomenclature[I]{\(m\in \mathcal{M}\)}{Index of regions}
\nomenclature[I]{\(n \in\mathcal{N}(m)\)}{Index of area that have renewable power transactions with area $m$}
\nomenclature[I]{\(t \in \mathcal{T}\)}{Index of time period}
\nomenclature[P]{\(\alpha_{m,0}/\alpha_{m,k}\)}{RPS target in region $m$ in base case/ non-base scenario $k$}
\nomenclature[P]{\(M\)}{Number of regions}
\nomenclature[P]{\(R^{\mathrm{u}}/R^{\mathrm{d}}\)}{Maximum available upward/downward reserve for conventional units.}
\nomenclature[P]{\(\epsilon_{k}\)}{Probability of $k^{th}$ scenario}
\nomenclature[P]{\(\bar{C}/\underline{C}\)}{Upward/Downward re-dispatch bid-in price of conventional units}
\nomenclature[P]{\(f\)}{Vector of transmission line capacities}
\nomenclature[P]{\(W_{t}\)}{Maximum available base-case renewable generation output at time $t$}
\nomenclature[P]{\(\Delta g^{U}/\Delta g^{D}\)}{Upward/Downward ramping rate limit of conventional units}
\nomenclature[P]{\(\phi^{\mathrm{d}}_{k,t}/\phi^{\mathrm{w}}_{k,t}/\phi^{\mathrm{g}}_{k,t}\)}{Load/RES/conventional units power deviation vectors at time $t$ in scenario $k$}
\nomenclature[P]{\({d}_{t}\)}{Base load at time $t$}
\nomenclature[P]{\(c_{\mathrm{g}}/c_{\mathrm{u}}/c_{\mathrm{d}}\)}{Vector of bid-in prices for energy and upward/downward reserves submitted by all conventional units}
\nomenclature[P]{\(S_{\mathrm{g}}/S_{\mathrm{w}}/S_{\mathrm{d}}\)}{Shift-factor matrix between lines and conventional units/renewable units/load}
\nomenclature[V]{\({g}_{t}/{w}_{t}\)}{Energy procured from conventional/renewable units at time $t$}
\nomenclature[V]{\(r_{t}^{\mathrm{u}, \mathrm{g}}/r_{t}^{\mathrm{d},\mathrm{g}}\)}{Conventional units upward/downward reserve at time $t$}
\nomenclature[V]{\(r_{t}^{\mathrm{u}, \mathrm{w}}/r_{t}^{\mathrm{d},\mathrm{w}}\)}{Renewable units upward/downward reserve at time $t$}
\nomenclature[V]{\(\delta \mathrm{g}_{k,t}^{+}/\delta \mathrm{g}_{k,t}^{-}\)}{Upward/Downward re-dispatch of conventional units at time $t$ in scenario $k$}
\nomenclature[V]{\(\delta \mathrm{w}_{k,t}^{+}/\delta \mathrm{w}_{k,t}^{-}\)}{Upward/Downward re-dispatch of RES at time $t$ in scenario $k$}
\printnomenclature[0.8in]  
\section{Introduction}
\IEEEPARstart{T}{he} carbon neutrality target promotes the integration of renewable energy. As the portion of renewable resources in the energy mix expands, the intermittent and uncertain nature of renewable energy supply adds to overall system flexibility requirements, making it challenging to accommodate RES in cost-oriented economic dispatch. As a result, system operators may sometimes have to curtail wind or solar generation~\cite{b1}. One government remedy for renewable curtailment is the Renewable Portfolio Standard (RPS), which mandates that a certain percentage of the electricity utilities sold come from renewable resources~\cite{RPS}. However, the implementation of RPS may intensify the flexibility shortage. As intermittent wind and solar power replace conventional units, the burden on conventional units providing ancillary services is becoming unprecedented. Emerging technology enables renewable resources themselves to act as reserve providers. However, whether renewable units should participate in reserve provision and how they participate are still open questions.\par
Several system operators have investigated the viability of utility-scale wind and solar power to provide spinning reserves. For instance, the Danish system operator Energinet conducted a pilot project, as documented in~\cite{Energinet}, while the California Independent System Operator (CAISO) performed a series of tests, detailed in~\cite{CAISO}. These studies demonstrated that wind turbines could provide downward reserves by adjusting their rotational speed or altering the pitch angle of their blades. Also, adjusting the turbine from operating at its maximum power point can enable the provision of upward reserves. Similarly, solar units can contribute to spinning reserves through delta control, as outlined in~\cite{b16}. However, renewable resources typically result in a loss of revenue during the provision of reserves. Renewable generators need a financial incentive to follow the system operator dispatch for reserve provision.

In academia, prior studies show that enabling reserve provision from renewable resources can relieve system flexibility~\cite{b18}-\hspace*{-5px}\cite{b19}, reduce total operation cost~\cite{b20}, and promote renewable penetration~\cite{b21}. Considering that the outputs from renewable resources are more unpredictable, a clear distinction is made between reserves procured from renewable resources and those from conventional generators. Many studies have noticed this distinction, utilizing various stochastic renewable reserve capacity modeling techniques, such as scenario-based stochastic optimization~\cite{Y}-\hspace*{-5px}\cite{b5}, CVaR-based optimization~\cite{b6}-\hspace*{-5px}\cite{b7}, and chance-constrained optimization~\cite{b8}-\hspace*{-5px}\cite{b10}. However, the associated pricing and market properties in these existing efforts have not been properly addressed, yet in~\cite{b15}, renewable producers' capability to provide flexibility is modeled as the re-dispatch quantity in the energy-only market. Also, cost recovery by scenario for renewable producers has been established. 

Furthermore, renewable resources are required to pay for the uncertainties they introduce endogenously. On the one hand, mandating increased integration of renewable energy sources through renewable portfolio standards can potentially lead to shortages in operational flexibility. On the other hand, excessive renewable resources could also pull down energy clearing prices. This may make it difficult for renewable resources to cover the integration costs associated with their power deviation. In light of this, a proposal for allocating reserve costs with Renewable Portfolio Standards implemented is put forth in~\cite{b14}. Meanwhile, spinning reserves are also vital in guarding against emergency incidents, such as generator outages. Conventional generators are required to pay the penalties or fines to the system operator in an outage, as discussed in ~\cite{b44}. Given these considerations, whether reserves procured from renewable and conventional units at the same bus are homogeneous is worth questioning. Can they settle at the same price? Additionally, the ability of the settlement to guarantee cost recovery is a crucial aspect that requires further investigation.\par
This paper considers the procurement, pricing and properties of integrating RESs as reserve providers. Compared to the existing literature, the main contributions of this paper are summarized below:
\begin{enumerate}
\item This paper develops a scenario-oriented energy-reserve joint optimal dispatch model. The optimal dispatch model explicitly incorporates regional Renewable Portfolio Standards, reserve provisions from renewable units, and outages in conventional units. Based on the proposed model, marginal prices of energy, reserve and power deviation are derived. The associated settlement for generators, structured in four parts, is developed: energy, reserve, and power deviation payments in the ex-ante stage and re-dispatch payments in the ex-post stage. 
\item Under the proposed settlement process, we show that thermal and renewable generators are equivalent when uncertainty and flexibility are considered as a whole. Units on the same bus face different costs due to uncertainties such as outages and power deviations, and they earn different credits for their ability to provide reserves and respond to re-dispatch. However, when considering these costs and credits as a whole, they are equal to the net re-dispatch settled at the uniform price under mild conditions, highlighting the inherent homogeneity of energy, regardless of generation source.
\item For any price-taking RES, providing reserve and possibly less energy is more profitable in the proposed mechanism, compared to generating as much as possible. When RES provide reserve capacity, they are able to partially offset net load power deviations. The opportunity cost to RES of providing reserve capacity rather than generating energy is compensated by nodal-uniform reserve clearing prices. 
\end{enumerate}\par
The remainder of this paper is organized as follows. Section II introduces the RPS-constrained energy and reserve joint optimal dispatch model. Section III provides the associated pricing and settlement process. Market properties are then discussed in Section IV. Section V conducts a series of sensitivity analyses, presenting numerical results contrasted with scenarios in which RESs exclusively supply energy. Section VI addresses concluding remarks. The associate proofs are left in the Appendix.
\vspace{-1em}\section{Model Formulation}
In this section, we consider a RPS-constrained multi-period optimal dispatch model where load and renewable power deviations from the base-case value and generator outages are modelled through a finite number of non-base scenarios with given probabilities. To formulate the regional RPS problem, we begin by distinguishing between two types of renewable output: the actual renewable energy production $w_{m,t}$ in region $m$ and the total amount $q_{m,t}$ is utilized to fulfil the RPS target in region $m$. Suppose there are $M$ regions in the entire system. For time interval $t$, renewable output $w_{t}$ and load power $d_{t}$ for the entire system are formulated as
\begin{align}
    \hspace{-2em}  w_{t}=\left({w}_{1,t}, \ldots, {w}_{M,t}\right) \in \mathbbm{W}:& =\mathbbm{W}_{1} \times \ldots \times \mathbbm{W}_{M}, \nonumber \\
    {d_{t}}=\left({d}_{1,t}, \ldots, {d}_{M,t}\right) \in \mathbbm{D}:& =\mathbbm{D}_{1} \times \ldots \times \mathbbm{D}_{M}.
\end{align} \par
The dimension of $w_{m,t}$ corresponds to the number of renewable units in region $m$. For each area $m$, the vector of renewable power output $w_{m,t}$ is decomposed into two components: the self-production part $w_{mm,t}$, and the aggregate net export $\sum\max \left(w_{mn, t}, 0\right)$, 
\begin{align}
 & \hspace{-1.2em} w_{m,t}=(w_{mm, t}+\sum_{n \neq m} \max \left(w_{mn, t}, 0\right)) \in \mathbbm{W}_{m}, \forall m\in\mathcal{M},
\end{align}
where $w_{mn,t}$ represents net exports from area $m$ to area $n$ as a positive value, and net imports as a negative value. The trading volume sold by area $m$ to area $n$ is identical to the volume purchased by area $n$ from area $m$, denoted as $w_{mn,t}=-w_{nm,t}$.
Subsequently, the scalar $q_{m,t}$ represents the total amount of self-generated electricity combined with the aggregate of net renewable energy imports:
\begin{align}
   \hspace{-1.2em} q_{m,t}=(\mathbf{1}^{\top}w_{mm,t}+ \sum_{n \in \mathcal{N}(m)} \mathbf{1}^{\top}\max \left(w_{nm, t}, 0\right)), \forall m\in\mathcal{M},
\end{align}\par
Additionally, we make some assumptions to formulate the optimal dispatch model. 
\begin{enumerate}
    \item An energy-reserve co-optimization model with regional renewable portfolio standard constraints is developed under a shift factor-based DC optimal power flow model. 
    \item Linear bid-in cost functions are considered for energy, reserve, and re-dispatch from conventional generators.
    \item Renewable producers offer their energy, reserve, and re-dispatch production at price zero\footnote{\label{1}In light of the limited practice of RES participating in the ancillary services markets, we follow National Renewable Energy Laboratory (NREL)'s official documents \cite{b23}-\cite{b24} and set reserve bid-in costs related to RES as zero.}.
    \item Each generator's lower bound is assumed to be zero.
\end{enumerate}

Based on the above formulations and assumptions, we present our optimal dispatch model in (\ref{eq:4}), with multipliers of all constraints listed on the left.\footnote{\label{2} Throughout the paper, we adopt "-" for lower-bound inequality constraints in the Lagrangian so all dual variables associated with the lower-bound inequality constraints are non-negative.} 
\begin{subequations}\label{eq:4}
\begin{align}
&\underset{x}{\min}F=\sum_{t\in T}\left(\begin{array}{ll}c_{\mathrm{g}}^{\top} g_{t}+c_{\mathrm{u}}^{\top} r_{t}^{\mathrm{u},\mathrm{g}}+c_{\mathrm{d}}^{\top} r_{t}^{\mathrm{d}, \mathrm{g}}+\\
\sum_{k\in\mathcal{K}} \epsilon_{k}(\overline{C}^{\top} \delta \mathrm{g}_{k,t}^{+}-\underline{C}^{\top}\delta\mathrm{g}_{k,t}^{-})\end{array}\right) , \\
&\text{subject to:}\text { for all } t \in \mathcal{T}: \nonumber\\
&\left(\lambda_{t}, \mu_{t}\right): \mathbf{1}^{\top} {g}_{t}+\mathbf{1}^{\top} {w}_{t} = \mathbf{1}^{\top} d_{t}, S_{g}g_{t}+S_{w}{w}_{t}-S_{d}d_{t} \leq f, \\
& \left(\underline{\iota ^{U}_{t}}, \overline{\iota ^{U}_{t}}, \underline{\iota ^{D}_{t}}, \overline{\iota ^{D}_{t}}\right): 0 \leq r_{t}^{\mathrm{u}, \mathrm{w}} \leq \bar{W}_{t}-{w}_{t}, 0 \leq r_{t}^{\mathrm{d}, \mathrm{w}} \leq {w}_{t}, \\
&\left(\underline{\imath_{t}}, \bar{\imath}_{t}\right):  \underline{G}+r_{t}^{\mathrm{d}, \mathrm{g}} \leq {g}_{t},{g}_{t}+r_{t}^{\mathrm{u}, \mathrm{g}} \leq \bar{G},\\
&\left(\underline{\ell^{U}_{t}}, \overline{\ell^{U}_{t}}, \underline{\ell^{D}_{t}}, \overline{\ell^{D}_{t}}\right): 0 \leq r_{t}^{\mathrm{u}, \mathrm{g}} \leq R^{\mathrm{u}},0 \leq r_{t}^{\mathrm{d}, \mathrm{g}} \leq R^{\mathrm{d}},\\
&\left(\gamma_{t}^{D},\gamma_{t}^{U}\right):
r_{t}^{\mathrm{d}, \mathrm{g}}-\Delta g^{D} \leq g_{t+1}-g_{t}\leq \Delta g^{U}-r_{t}^{\mathrm{u}, \mathrm{g}},\\
&\left(\nu_{m,0}\right): \sum_{t\in T}\left(\alpha_{m,0}^{\top}d_{m,t}\right)\leq \sum_{t\in T}q_{m,t},\forall m\in \mathcal{M},\\
&\text { for all } k \in \mathcal{K}: \nonumber\\
&\lambda_{k,t}: \mathbf{1}^{\top}\left({g}_{t}+\delta \mathrm{g}_{k,t}^{+}-\delta \mathrm{g}_{k,t}^{-}\right) \nonumber \\ 
& \quad+\mathbf{1}^{\top}\left({w}_{t}+\delta \mathrm{w}_{k,t}^{+}-\delta \mathrm{w}_{k,t}^{-}\right)=\mathbf{1}^{\top}\left(d_{t}+\phi_{k,t}^{\mathrm{d}}\right),\\
&\mu_{k,t}: S_{\mathrm{g}}\left({g}_{t}+\delta \mathrm{g}_{k,t}^{+}-\delta \mathrm{g}_{k,t}^{-}\right)\nonumber \\
& \quad+S_{\mathrm{w}}\left({w}_{t}+\delta \mathrm{w}_{k,t}^{+}-\delta \mathrm{w}_{k,t}^{-}\right) -S_{\mathrm{d}}^{\top}\left(d_{t}+\phi_{k,t}^{\mathrm{d}}\right)\leq f, \\
&\left(\underline{\eta_{k,t}}, \overline{\eta_{k,t}}\right): 0 \leq \delta \mathrm{g}_{k,t}^{+} \leq (\mathbf{I}-\mathbf{X}_{k}) r_{t}^{\mathrm{u}, \mathrm{g}}, \\
&\left(\underline{\beta_{k,t}}, \overline{\beta_{k,t}}\right): 0 \leq \delta \mathrm{g}_{k,t}^{-}-\phi_{k,t}^{\mathrm{g}} \leq (\mathbf{I}-\mathbf{X}_{k})r_{t}^{\mathrm{d}, \mathrm{g}}, \\
&\left(\underline{\tau_{k,t}} \overline{\tau_{k,t}}\right): 0 \leq \delta \mathrm{w}_{k,t}^{+} \leq r_{t}^{\mathrm{u}, \mathrm{w}}+\phi_{k,t}^{\mathrm{w},+}, \\
&\left(\underline{\zeta_{k,t}}, \overline{\zeta_{k,t}}\right): 0 \leq\delta \mathrm{w}_{k,t}^{-}-\phi_{k,t}^{\mathrm{w},-} \leq r_{t}^{\mathrm{d}, \mathrm{w}}, 
\end{align}
\begin{align}
&\left(\nu_{m, k}\right): \sum_{t \in T} \alpha_{m,k}^{\top}\left(d_{m,t}+\phi_{m,k,t}^{\mathrm{d}}\right) \leq \sum_{t\in T}\left (\mathbf{1}^{\top}\delta {w}_{m,k,t}+q_{m, t}\right),\nonumber
\\ & \quad \quad \quad \forall m \in \mathcal{M},
\end{align}
\end{subequations}
where $x=\{{g}_{t},{w}_{t}, r_{t}^{\mathrm{u}, \mathrm{g}}, r_{t}^{\mathrm{d}, \mathrm{g}}, r_{t}^{\mathrm{d}, \mathrm{w}},r_{t}^{\mathrm{u}, \mathrm{w}},\delta \mathrm{g}_{k,t}^{+},\delta \mathrm{g}_{k,t}^{-},\delta \mathrm{w}_{k,t}^{+},$\quad$\delta \mathrm{w}_{k,t}^{-}\}$ is the set of decision variables. The objective function in (\ref{eq:4}a) aims to minimize the expected total cost which includes energy procurement $\sum_t c_{\mathrm{g}}^{\top} g_{t}$, upward and downward reserve procurement $\sum_t c_{\mathrm{u}}^{\top} r_{t}^{\mathrm{u},\mathrm{g}}+\sum_t c_{\mathrm{d}}^{\top} r_{t}^{\mathrm{d}, \mathrm{g}}$, plus the expectation of re-dispatch cost $\sum_t  \sum_k \left(\epsilon_{k}\left(\overline{C}^{\top} \delta \mathrm{g}_{k, t}^{+}-\underline{C}^{\top} \delta \mathrm{g}_{k, t}^{-}\right)\right)$. Constraint (\ref{eq:4}b) denotes energy balancing and transmission capacity constraints for the base case. Constraint~(\ref{eq:4}c) represents capacity limits for renewable units' output and reserve. Constraint (\ref{eq:4}d) denotes the lower and upper bounds for conventional units' power generation and reserve. In addition, owing to inertia, conventional units are enforced with reserve capability limits (\ref{eq:4}e). Constraint (\ref{eq:4}f) represents ramping limits between two adjacent time intervals to guarantee conventional units' reserve deliverability. Take upward reserve as an example, the sum of upward ramping $g_{t+1}$- $g_{t}$ from period $t$ to period $t+1$ and procured upward reserve $r_{U,t}$ at period $t$ cannot exceed the maximum upward ramping rate $\Delta g^{U}$ between sequential periods. Such formulation can also be found in\cite{b3}. Since the system is constrained by flexibility conditions, e.g., reserve requirements and ramping limits, the system will not schedule renewable generation at full capacities. Constraint (\ref{eq:4}g) denotes regional RPS for base-case production in a day. Constraints (\ref{eq:4}h)~-~(\ref{eq:4}n) are scenario-dependent constraints for non-base case $k$. Constraints (\ref{eq:4}h)-(\ref{eq:4}i) represent power balance equation and line capacities limits for non-base case $k$, respectively. Constraints (\ref{eq:4}j)-(\ref{eq:4}k) indicate that available reserve capacities procured from conventional generators will be their maximum re-dispatch among all non-base scenarios. The diagonal matrix denoted by $\mathbf{X_k}$ describes the state of conventional generators in scenario $k$. If generator $i$ is shut down in scenario $k$, the $i^{th}$ diagonal element of $\mathbf{X}_{k}$ is set to one, and it is set to zero otherwise. Subsequently, the power deviation vector for conventional generators in scenario $k$ can be calculated as $\phi_{k,t}^{\mathrm{g}} = \mathbf{X}_{k}g_{t}$. The subtraction of the diagonal matrix from the identity matrix $\mathbf{I}-\mathbf{X}_{k}$ represents the state of available reserve capacities. Constraints (\ref{eq:4}l)\text{-}(\ref{eq:4}m) enforce RES re-dispatch in scenario $k$ shall not exceed base-case scheduled reserve and power deviation in scenario $k$. We define two additional non-negative auxiliary variables\footnote{\label{3}Note that load and renewable resources cannot be curtailed below zero in any non-base scenario. To consider this, we only need to add two additional constraints $d_{t}+\phi_{k,t}^{\mathrm{d}} \geq 0$ and $\phi_{k,t}^{\mathrm{w},-}\leq \bar W_t$ and two associated multipliers into model (\ref{eq:4}), and all the qualitative analyses will still hold.}, positive imbalance $\phi_{k,t}^{\mathrm{w},+}$, and negative imbalance $\phi_{k,t}^{\mathrm{w},-}$. Constraint (\ref{eq:4}n) represents non-base RPS constraints. Vector $\delta {w}_{m,k,t}$ denotes the net re-dispatch of renewable units in area $m$ for scenario $k$ at time $t$.\par
Note that the proposed scenario-based optimal dispatch model (\ref{eq:4}) is a standard linear programming problem, and it can be efficiently solved by implementing decomposition techniques such as benders decomposition\cite{b4}, critical region exploration\cite{b25} or progressive hedging algorithm\cite{b26}. However, solution methods are outside the scope of this paper. Compared to the model in \cite{b3}, the proposed dispatch model includes regional Renewable Portfolio Standards, renewable reserve capacities, and conventional unit outages. These modifications may affect the price uniformity and other market properties, which will be discussed in the following sections.
\section{Pricing and Settlement}
This section presents the pricing scheme and associated market settlement process. In the ex-ante stage, prices for energy, reserve and power deviation are derived through marginal pricing. In the ex-post stage, power re-dispatch is settled as pay-as-bid. For notation clarity, we use scalar energy price $\pi(i)$, upward and downward prices $\pi^u(i)$ and $\pi^d(i)$, and power deviation price $\pi^\phi_k(i)$ in scenario $k$ for each market participant $i$. Furthermore, we use superscripts $g$, $w$, and $d$ to denote the prices applicable to conventional units, renewable units, and loads, respectively. We use the notation $\mathcal{L}$ to represent the Lagrangian of the proposed optimal dispatch model (\ref{eq:4}).
\vspace{-0.5em}\subsection{Price Formation for Conventional Units}
In this subsection, we begin by defining the parametric optimal dispatch model. Consider any conventional generator $i$ at time $t$, by fixing variables at optimal value $g_{i,t}^{*}, (r_{i,t}^{\mathrm{u},\mathrm{g}})^{*},(r_{i,t}^{\mathrm{d}, \mathrm{g}})^{*}$, we exclude the bid-in cost of generator $i$ in the objective function (\ref{eq:4}a) to form ${F_{-it}}$. We also exclude resource constraints associated unit $i$ in model (\ref{eq:4}) to form a partial Lagrangian function ${\mathcal{L}_{-it}}$ and a parameterized optimal dispatch model $\underset{x_{-i}\in \mathcal{X_{-i}}}{\min}F$,
where $x_{-i}$ denotes the decision variables except for $g_{i,t},r_{i,t}^{\mathrm{u},\mathrm{g}},r_{i,t}^{\mathrm{d},\mathrm{g}}$, and ${X_{-i}}$ denotes constraints (\ref{eq:4}b)-(\ref{eq:4}n) except for resource-level constraints of generator $i$. According to the KKT conditions, the solution to model (\ref{eq:4}) is always optimal for the parameterized model.
\par
Subsequently, for any conventional unit $i$ at bus $b_{i}$, at the optimal solution $g_{i,t}^{*}$, energy marginal price is defined as the marginal benefit of one unit of $g_{i,t}^{*}$ to the rest of the market:
\begin{align}\label{eq:5}
       \pi_{t}^{g}(i)&=- \frac{\partial F_{-it}^{*}}{\partial g_{i,t}^{*}}=- \frac{\partial \mathcal{L}_{-it}}{\partial g_{i,t}^{*}}\nonumber \\&=\lambda_{t}^{*}-S_{\mathrm{g}}\left(:,b_{i}\right)^{\top} \mu_{t}^{*}+\sum_{k \in \mathcal{K}}\left(\lambda_{k, t}^{*}-S_{\mathrm{g}}\left(:,b_{i}\right)^{\top} \mu_{k, t}^{*}\right)\nonumber \\&=\pi_{0,t}^{g}(i)+\sum_{k \in \mathcal{K}}\pi_{k,t}^{g}(i), 
\end{align}

 Likewise, by the envelope theorem, we derive upward and downward reserve marginal prices respectively as follows:
  \vspace{-0.5em}\begin{align}
&\hspace{-0.5em}\pi_{t}^{\mathrm{u}, \mathrm{g}}(i)=-\frac{\partial \mathcal{L}_{-it}}{\partial (r_{i,t}^{\mathrm{u},\mathrm{g}})^{*}}=\sum_{k \in \mathcal{K},i \notin \Omega_{K}} \overline{\eta_{k, t}}(i)=\sum_{k \in \mathcal{K},i \notin \Omega_{K}} \pi_{k,t}^{\mathrm{u}, \mathrm{g}}(i),\\
&\hspace{-0.5em}\pi_{t}^{\mathrm{d}, \mathrm{g}}(i)=-\frac{\partial \mathcal{L}_{-it}}{\partial (r_{i,t}^{\mathrm{d},\mathrm{g}})^{*}}=\sum_{k \in \mathcal{K},i \notin \Omega_{K}} \overline{\beta_{k, t}}(i)=\sum_{k \in \mathcal{K},i \notin \Omega_{K}} \pi_{k,t}^{\mathrm{d}, \mathrm{g}}(i),
 \end{align}
where the set $\Omega_{K}$ represents the collection of generators that are experiencing an outage in scenario $k$. \par
Now, we derived the power deviation prices in scenario $k$ at time $t$. For a conventional unit, there is only downward power deviation due to generator outage. The pricing of downward power deviation is based on the marginal contribution of $\phi_{k,t}^{\mathrm{g}}(i)$ to the optimal total cost $F^*$, which following the principle of {\em cost allocation based on cost causation}\cite{H}. By applying the envelope theorem, the price of downward power deviation in scenario $k$ at time $t$ is defined as: 
\hspace{-1em}\begin{align}\label{eq:8}
    \pi^{\phi,g}_{k,t}(i) =\frac{\partial F^*}{\partial \phi_{k,t}^{\mathrm{g}}(i)}=\frac{\partial \mathcal{L}}{\partial \phi_{k,t}^{\mathrm{g}}(i)}=-\overline{\beta_{k, t}}(i)+\underline{\beta_{k, t}}(i),\forall i \in \Omega_{K}.
\end{align}
 where the multipliers $\overline{\beta_{k, t}}(i)$ and $\underline{\beta_{k, t}}(i)$ are dual variables associated with the upper and lower bounds of downward re-dispatch constraint (\ref{eq:4}k) for outage conventional unit $i$ at scenario $k$. Setting the partial derivative of the Lagrangian with respect to the variable $\delta \mathrm{g}_{k, t}^{-}(i)^{*}$ equal to zero, we obtain the condition for optimality:
\begin{align}\label{eq:9}
    \hspace{-1em}\frac{\partial \mathcal{L}}{\partial\delta \mathrm{g}_{k, t}^{-}(i)^{*}}=-\epsilon_{k} \underline{C}(i)+\overline{\beta_{k, t}}(i)-\underline{\beta_{k, t}}(i)+\pi_{k, t}^{\mathrm{g}}(i)= 0.
\end{align}\par
Substituting (\ref{eq:9}) into the right-hand side of (\ref{eq:8}), we have
\begin{align}
 \pi^{\phi,g}_{k,t}(i)=\pi_{k,t}^{g}(i)-\epsilon_{k} \underline{C}(i),\forall i \in \Omega_{K},
\end{align}
which is equivalent to the difference between contingency scenario $k's$ contribution to the energy price and the expected re-dispatch pay-back.
\vspace{-1em}\subsection{Price Formation for Renewable Units}
Similarly, by converting variables to parameters fixed at the optimal values and removing the related resource-level constraints, we establish the parametric optimal dispatch model $\underset{x_{-j}\in \mathcal{X_{-j}}}{\min}F$ for any renewable producer $j$. Let $b_{j}$ be the bus index and $m_{j}$ be the area index where renewable power plant $j$ is located. Based on the envelope theorem, we derive the energy marginal price for renewable unit $j$ as:
\begin{align}\label{eq:11}
     \pi_{t}^{w}(j)=&-\frac{\partial \mathcal{L}_{-jt}}{\partial w_{j,t}^{*}}
     =\lambda_{t}^{*}-S_{\mathrm{w}}\left(:,b_{j}\right)^{\top} \mu_{t}^{*}+\nu^{*}(m_{j})\nonumber \\+&\sum_{k \in \mathcal{K}}\left(\lambda_{k, t}^{*}-S_{\mathrm{w}}\left(:,b_{j}\right)^{\top} \mu_{k, t}^{*}+\nu_{k}^{*}(m_{j})\right)\nonumber \\
     =&\pi_{0,t}^{w}(j)+\sum_{k \in \mathcal{K}}\pi_{k,t}^{w}(j),
 \end{align}
where $\pi_{t}^{w}(j)$ is the sum of the base component $\pi_{0,t}^{w}(j)$ and non-base component $\pi_{k,t}^{w}(j)$. Each component has a locational-uniform energy part and a regional-uniform RPS part. Dual variables $\nu^{*}$ and $\nu_{k}^{*}$ associated with RPS constraints (\ref{eq:4}g) and (\ref{eq:4}n) can be interpreted as the incremental system cost to accommodate another unit of RES output. \par
Likewise, we compute downward reserve marginal price $\pi_{t}^{\mathrm{d}, \mathrm{w}}(j)$ and upward reserve marginal price $\pi_{t}^{\mathrm{u}, \mathrm{w}}(j)$ for renewable unit $j$ at bus $b_{j}$ respectively as follows:
\begin{align}
&\hspace{-1em}\pi_{t}^{\mathrm{d}, \mathrm{w}}(j)=-\frac{\partial \mathcal{L}_{-jt}}{\partial (r_{j,t}^{\mathrm{d},\mathrm{w}})^{*}}=\sum_{k \in \mathcal{K}} \overline{\zeta_{k, t}}(j)=\sum_{k \in \mathcal{K}} \pi_{k,t}^{\mathrm{d}, \mathrm{w}}(j),  \\ &\hspace{-1em}\pi_{t}^{\mathrm{u}, \mathrm{w}}(j)=-\frac{\partial \mathcal{L}_{-jt}}{\partial \left(r_{j,t}^{\mathrm{u},\mathrm{w}}\right)^{*}}(j)=\sum_{k \in \mathcal{K}} \overline{\tau_{k, t}}(j)=\sum_{k \in \mathcal{K}} \pi_{k,t}^{\mathrm{u}, \mathrm{w}}(j).
 \end{align}\par
 Hereafter, we derive upward power deviation price at time $t$ in scenario $k$:
 \begin{align}
 \pi_{k,t}^{\phi,w,+}(i)= \frac{\partial \mathcal{L}}{\partial \phi_{k,t}^{\mathrm{w,+}}(j)}=\overline{\tau_{k, t}}(j)=\pi_{k,t}^{\mathrm{u}, \mathrm{w}}(j),
\end{align}\par
 Similarly, downward power deviation price at scenario $k$:
  \begin{align}
 \pi_{k,t}^{\phi,w,-}(i)= \frac{\partial \mathcal{L}}{\partial \phi_{k,t}^{\mathrm{w,-}}(j)}=-\overline{\zeta_{k, t}}(j)+\underline{\zeta_{k, t}}(j).
\end{align}\par
Note that the first-order optimally condition with respect to $\delta \mathrm{w}_{k, t}^{-}(i)$ implies,
\begin{align}\label{eq:16}
&\frac{\partial \mathcal{L}}{\partial(\delta \mathrm{w}_{k, t}^{-}(i))^{*}}=\overline{\zeta_{k, t}}(i)-\underline{\zeta_{k, t}}(i)+\pi_{k,t}^{w}(i)= 0.
\end{align}\par
Consequently, each renewable unit $i$ in scenario $k$ at time $t$ is charged for its production deficit $-\phi_{k,t}^{\mathrm{w},-}(i)$ at a price $\pi_{k,t}^{w}(i)$ and is paid for its production surplus $\phi_{k,t}^{\mathrm{w},+}(i)$ at a price $\pi_{k,t}^{\mathrm{u}, \mathrm{w}}(i)$. 
\vspace{-1em}\subsection{Price Formation for Loads}
The pricing approach treats load as an inflexible negative renewable resource, including only energy and power deviation pricing. Correspondingly, by taking the partial derivative of the Lagrangian function $\mathcal{L}$ over $d_{l,t}$, we can derive the energy marginal price for load demand $l$ at bus $b_{l}$ in area $m_{l}$ as
\begin{align}\label{eq:17}
    \pi_{t}^{d}(l)=&\frac{\partial F^{*}}{\partial d_{l,t}}= \frac{\partial \mathcal{L}}{\partial d_{l,t}}\nonumber \\ 
    =&\lambda_{t}^{*}-S_{\mathrm{d}}\left(:,b_{l}\right)^{\top} \mu_{t}^{*}+\alpha_{m,0}\cdot\nu^{*}(m_{l})\nonumber \\&+\sum_{k \in \mathcal{K}}\left(\lambda_{k, t}^{*}-S_{\mathrm{d}}\left(:,b_{l}\right)^{\top} \mu_{k, t}^{*}+\alpha_{m,k}\cdot\nu_{k}^{*}(m_{l})\right)\nonumber \\
    =&\pi_{0,t}^{d}(l)+\sum_{k \in \mathcal{K}} \pi_{k,t}^{d}(l),
\end{align}
which is the sum of the base component $\pi_{0,t}^{d}(l)$ and non-base component $\pi_{k,t}^{d}(l)$. Each component is made up of two parts: (i) the locational-uniform energy component which is similar to the conventional unit's energy price proposed in (\ref{eq:4}), (ii) the RPS-related part $\alpha_{m}\nu^{*}(m_{l})$ when procuring one extra MW of renewable generation. For the loads at the same bus, the two components of pricing are consistent, thereby ensuring uniform energy pricing for loads at the same bus. However, with the price formation in equations (\ref{eq:5}), (\ref{eq:11}), and (\ref{eq:17}), conventional generators, RESs, and loads at the same bus may receive different marginal energy prices when RPS constraints are active.  \par
Similarly, power deviation price for load $l$ at time $t$ in scenario $k$ is calculated as, 
\begin{align}
    \pi_{k,t}^{d}(l)=&\frac{\partial\mathcal{L}}{\partial \phi_{k,t}^{\mathrm{d}}(l)}= \pi_{k,t}^{d}(l).
\end{align}
\vspace{-2em} \subsection{Market Settlement}
In this subsection, the general payment expression for each resource is established, with the previously defined price notations simplified by omitting the superscripts. We denote energy, re-dispatch, upward reserve, downward reserve, and power deviation as $x_t$, $\delta x_t$, $r^u_t$, $r^d_t$, and $\Phi_t$, respectively. Based on our observations from the previous formation of prices, we can deduce that the payment $\Gamma_i$ for any generator $i$ is comprised of four main components: 
 \begin{align}\label{eq:19}
\Gamma_i=\mathcal{E}_i+\mathcal{R}_i+\chi_i+\mathcal{D}_{k,i}
 \end{align}
 where the energy credit $\mathcal{E}_i$, the reserve credit $\mathcal{R}_i$, and the power deviation payment $\chi_i$, are settled in the ex-ante stage. The corresponding formulations are presented as follows:\par
Each generator $i$ receives payment from the system operator for its scheduled power $x_{i,t}$ at the corresponding price $\pi_{t}(i)$.
\begin{align}\label{eq:20}
    &\mathcal{E}_i=\sum_{t\in \mathcal{T}}\pi_{t}(i)x_{i,t}.
\end{align}\par
Each generator $i$ receives payment from the system operator for its scheduled upward reserve $r_{i,t}^{\mathrm{u}}$ at the price $\pi_{t}^{\mathrm{u}}(i)$ and for its scheduled downward reserve $r_{i,t}^{\mathrm{d}}$ at the price $\pi_{t}^{\mathrm{d}}(i)$. 
\begin{align}
    &\mathcal{R}_i=\sum_{t\in \mathcal{T}}\left(\pi_{t}^{\mathrm{d}}(i) r_{i,t}^{\mathrm{d}}+\pi_{t}^{\mathrm{u}}(i) r_{i,t}^{\mathrm{u}}\right).
\end{align}\par
Each generator $i$, when experiencing a power deviation in scenario $k$ from the base case, receives payment for its power surplus $\Phi^{+}_{k,t}(i)$ at the price $\pi_{k,t}^{\phi,+}(i)$, and incurs a cost for its power deficit $-\Phi^{-}_{k,t}(i)$ at the price $\pi_{k,t}^{\phi,-}(i)$. 
\begin{align}\label{eq:22}
    \chi_i&=\sum_{t\in \mathcal{T}}\sum_{k\in \mathcal{K}}\pi_{k,t}^{\phi}(i)\Phi_{k,t}(i)\nonumber\\
    &=\sum_{t\in \mathcal{T}}\sum_{k\in \mathcal{K}}(\pi_{k,t}^{\phi,+}(i)\Phi^{+}_{k,t}(i)-\pi_{k,t}^{\phi,-}(i)\Phi^{-}_{k,t}(i))
\end{align}\par
When generator $i's$ power deviation contributes to mitigating the net load imbalance, $\chi_{i}\geq 0$, indicating that producer $i$ receives credit for its deviation. Conversely, if the deviation exacerbates the imbalance, $\chi_{i}$ will be negative, representing a cost to the producer. \par
In the ex-post stage, with the realization of scenario $k$, power re-dispatch quantities are settled at their bid-in costs:
\begin{align}
\mathcal{D}_{k,i} = \sum_{t\in \mathcal{T}}\left(\bar{C}(i) \delta \mathrm{x}_{k,t}^{+}(i)-\underline{C}(i)  \delta \mathrm{x}_{k,t}^{-}(i)\right).
\end{align}\par
For load $i$, it pays system operator energy payment $\mathcal{E}^d_i$ and power deviation payment $\chi^d_i$ at the ex-ante stage. These payments are calculated based on quantities and prices outlined in equations (\ref{eq:20}) and (\ref{eq:22}).
So far, we've established the pricing and market settlement process. The properties of the proposed market settlement will be summarized in the next section. 
\section{Market Properties}
This section investigates several properties, including locational uniform pricing for renewable reserve, the equivalence of thermal and renewable units when considering uncertainty and flexibility as a whole, individual rationality and cost recovery for generators, and revenue adequacy for the system operator. \par
First, we develop uniform renewable reserve pricing, as opposed to discriminative reserve prices from conventional units in \cite{b3}. 
\begin{theorem}[Uniform Pricing for Renewable Reserve]\label{th:1}
 Consider any two renewable units $i$ and $j$ at the same bus. Under assumption (3) and that reserve prices are derived in (8)-(9). Renewable units $i$ and $j$ have the same reserve price. i.e., $\pi_{t}^{\mathrm{u}, \mathrm{w}}(i)=\pi_{t}^{\mathrm{u}, \mathrm{w}}(j)$, and $\pi_{t}^{\mathrm{d}, \mathrm{w}}(i)=\pi_{t}^{\mathrm{d}, \mathrm{w}}(j)$.\end{theorem} \par
 The detailed proof is provided in Appendix A. The fractional energy price for generators in scenario $k$ is determined by the fractional reserve price in scenario $k$ and the expected re-dispatch cost, where the maximum re-dispatch quantity is equal to the available reserve capacity. Renewable generators are assumed to have re-dispatch costs of zero, resulting in uniform reserve prices. In contrast, thermal generators at the same bus may have distinct re-dispatch costs and outage probabilities, leading to discriminative reserve prices under the proposed pricing scheme. We consider the priority of reserve as follows:
\vspace{-0.5em}\begin{Corollary}
  Assume that re-dispatch costs of thermal units $\overline{C}$ and $\underline{C}$ are positive. Consider any renewable unit $j$ and conventional generator $i$ at the same bus, the upward reserve price of renewable unit $j$ is higher than that of conventional unit $i$, and renewable unit $j$ has a lower downward reserve price than generator $i$. i.e., $\pi_{t}^{\mathrm{u}, \mathrm{w}}(j)\geq\pi_{t}^{\mathrm{u}, \mathrm{g}}(i), \pi_{t}^{\mathrm{d}, \mathrm{w}}(j)\leq \pi_{t}^{\mathrm{d}, \mathrm{g}}(i)$.
\end{Corollary} \par
\vspace{-0.5em}The detailed proof is given in Appendix B. RESs have higher opportunity costs than thermal units to provide upward reserve, as RESs usually get dispatched first due to their low marginal cost. Conventional generators will pay more back to the system operator when the downward reserve is activated than renewable producers due to downward re-dispatch prices. Prioritizing upward reserve for renewable units and downward reserve for thermal units promotes renewable energy integration. \par 
Next, we demonstrate the equivalence between thermal generators and renewable units within the proposed pricing mechanism. This involves considering their uncertainties, such as RES power deviation and potential contingencies in thermal units, along with their flexibilities. For unit $i$, the combined fractional contributions in the realized scenario $k$, derived from reserve, re-dispatch, and power deviation, are defined as follows:
\vspace{-0.5em}\begin{align}
\Pi_{k,t}^{D}(i) & = \pi_{k,t}^{\mathrm{d}}(i) r_{i,t}^{\mathrm{d}}-\epsilon_{k} \underline{C} \delta x_{k,t}^{-}(i)- \chi_{k,t}^{-}(i), \\
\Pi_{k,t}^{U}(i) & = \pi_{k,t}^{\mathrm{u}}(i) r_{i,t}^{\mathrm{u}}+\epsilon_{k} \bar{C} \delta x_{k,t}^{+}(i)+\chi_{k,t}^{+}(i),
\end{align}
where $\pi_{k,t}^{\mathrm{u}}(i) r_{i,t}^{\mathrm{u}}$ and $\pi_{k,t}^{\mathrm{d}}(i) r_{i,t}^{\mathrm{d}}$ represent the fractional upward and downward reserve revenue for generator $i$, respectively. $\chi_{k,t}^{+}(i)$ and $\chi_{k,t}^{-}(i)$ correspond to contribution of scenario $k$ to the upward and downward power deviation payments for generator $i$ at time $t$. Hereafter, we establish thermal and renewable units equivalence as follows:
\begin{theorem}[Equivalence of Diverse Generators]\label{th:2}
Under assumptions (1)-(4), consider any two units, say i and j, located at the same bus. For any realized scenario k, the combined contributions of uncertainties and flexibilities $\Pi_{k,t}^{U}+\Pi_{k,t}^{D}$ is equivalent to settled the net re-dispatch $\left(\delta x_{k,t}^{U}-\delta x_{k,t}^{D}\right)$ at the corresponding fractional energy price $\pi_{k,t}$, i.e, \begin{align}\label{eq:26}
\Pi_{k,t}^{U}(i)+\Pi_{k,t}^{D}(i) & = \pi_{k,t}\left(i\right)\left(\delta x_{k,t}^{U}(i)-\delta x_{k,t}^{D}(i)\right), 
\end{align}
\begin{align}
\Pi_{k,t}^{U}(j)+\Pi_{k,t}^{D}(j) & = \pi_{k,t}\left(j\right)\left(\delta x_{k,t}^{U}(j)-\delta x_{k,t}^{D}(j)\right). 
\end{align}\par
Assuming RPS shadow price $\nu_{k}^*$ of this bus to be zero\footnote{A non-zero $\nu_{k}$ occurs when consuming additional RES in scenario $k$ leads to an overall increase in expected total cost, implying that the increased flexibility cost surpasses the reduced generation cost. Currently, most 
countries have set RPSs at or below 30\%~\cite{RPS} which satisfies the assumption that $\nu_{k}^*=0$. Even with a green premium, $\pi_{k}(j)=\pi_{k}(i)+\nu_{k}$, the homogeneity of the energy attribute is not violated.}, for any thermal unit i and renewable unit j located at the same bus, they have the uniform fractional energy price component, i.e,
$\pi_{k,t}\left(i\right)  = \pi_{k,t}\left(j\right), \forall k \in {\mathcal{K}}.$
\end{theorem} \par
\vspace{-0.5em}The proof is given in Appendix B. Theorem 2 indicates that though thermal and renewable units at the same bus pay different amounts for uncertainty and earn different credits for flexibility. Their combination is uniform, which means the fundamental principle of energy homogeneity. Even considering renewable reserve and re-dispatch bid-in costs, this property is valid. \par
Then, we establish the property of individual rationality.  Let $\mathbb{P}_{i}$ represents the set of capacity limits for conventional unit $i's$ energy, reserve and re-dispatch, i.e., $i-th$ entry of constraints (\ref{eq:4}d)~-~(\ref{eq:4}f) and (\ref{eq:4}j)~-~(\ref{eq:4}k). Let $\mathbb{W}_{i,t}$ denotes the set of capacity limits for renewable unit $i's$ energy, reserve and re-dispatch at time $t$, i.e., $i-th$ entry of constraints ~(\ref{eq:4}c) and (\ref{eq:4}l)~-~(\ref{eq:4}m). Let $\Xi^{g}_{i}$ be the total bid-in cost of energy and reserve and re-dispatch in the realized scenario for conventional unit $i$.
\begin{align}\label{eq:28}
\Xi^{g}_{i}  =& \sum_{t \in T} (c_{\mathrm{g}}(i) {g}_{i,t}+c_{\mathrm{u}}(i) r_{i,t}^{\mathrm{u}, \mathrm{g}}+c_{\text {d}}(i) r_{i,t}^{\mathrm{d}, \mathrm{g}})+\mathcal{D}_{k,i}.
\end{align}\par
Based on the total revenue from energy, reserve, re-dispatch and power deviation for thermal unit $\Gamma^{g}$ and renewable unit $\Gamma^{w}$ in~(\ref{eq:19}) and total bid-in cost $\Xi^{g}$ in~(\ref{eq:28}), we delineate individual rationality as the following:
\begin{theorem}[Individual Rationality]
\label{th:3}
 Assume any unit $i$ is a price-taker settled at prices $\{\pi_{t}(i)^*$,$\pi_{k,t}^{\phi}(i)^*$,$\pi_{t}^{\mathrm{u}}(i)^*$,$\pi_{t}^{\mathrm{d}}(i)^*\}$ that are optimally solved from model~(\ref{eq:4}). Profit maximization of any producer $i$ gives the same optimal solutions as system cost minimization. Producer $i$ can maximize its expected profit at the system optimum, i.e.,
\begin{align}
&\left(g_{i}^{*},\left(r_{i}^{\mathrm{d}, \mathrm{g}}\right)^{*},\left(r_{i}^{\mathrm{u}, \mathrm{g}}\right)^{*}\right) = \underset{\mathbb{P}_{i}}{\operatorname{argmax}} \sum_{t \in T}(\Gamma^{g}_{i}-\Xi^{g}_{i}),\label{eq:29}\\
&\left(w_{i}^{*},\left(r_{i}^{\mathrm{d}, \mathrm{w}}\right)^{*},\left(r_{i}^{\mathrm{u}, \mathrm{w}}\right)^{*}\right) = \underset{\mathbb{W}_{i,t}}{\operatorname{argmax}} \sum_{t \in T}\Gamma^{w}_{i}.\label{eq:30}
\end{align}
\end{theorem}\par
  Note that the lagrangian of cost minimization model (\ref{eq:4}) can be decomposed to individual profit maximization problems is straightforward. Consequently, the KKTs of profit-maximization model~(\ref{eq:29}) and~(\ref{eq:30}) match with the part of the KKTs of cost-minimization model~(\ref{eq:4}). Equation (\ref{eq:30}) implies that no renewable unit $i$ receive better-off profit by deviating from the optimal dispatch.
   \begin{Corollary}Under assumptions (1)-(4), consider any price-taking renewable unit i that is incapable of influencing price signals $\{\pi(i)^*$, $\pi_{k}^{\phi}(i)^*$, $\pi^{\mathrm{u}}(i)^*$, $\pi^{\mathrm{d}}(i)^*\}$ from the system operator. With other units fixed at the optimal dispatch solved from model~(\ref{eq:4}), we have 
    \begin{align}
     \Gamma^{w}_{i},(w^*_{i},(r_{i}^{\mathrm{u}, \mathrm{w}})^{*},(r_{i}^{\mathrm{d}, \mathrm{w}})^{*}) \geq \Gamma^{w}_{i}(w_{i,max},0,0), 
  \end{align} 
   where $ w_{i,max} $ is the maximum energy output of unit $i$ within the feasible region $\mathcal{X}$ defined by constraints (\ref{eq:4}a)-(\ref{eq:4}n), i.e, $\forall w_{i} > w_{i,max}, (w_{i},0,0,{x_{-i}^*})\notin \mathcal{X}$.
  \end{Corollary}\par
  In other words, for RESs, providing reserve according to the dispatch results compared to generating as much as possible leads to better-off profits. That is to say, renewable units have incentives to provide reserve according to dispatch, which can partially offset the net load power deviation. This corollary can be proven by the fact that given the optimal prices, the optimal quantities solved from model~(\ref{eq:4}) are the solutions that maximize the objective function of~(\ref{eq:30}), as no other feasible solution with a better objective function exists. Moreover, cost recovery for generators is established as follows:
\vspace{-0.25em}\begin{theorem}[Cost Recovery]\label{th:4}
Under assumptions (1)-(4), by following the optimal dispatch solved by the model~(\ref{eq:4}), for (\ref{eq:29}), the total bid-in profit of any thermal unit i for supplying energy and reserve, and paying for the power deviation, is non-negative, i.e., $\Gamma^{g*}_{i} \geq\Xi^{g}_{i}$. The same holds true for renewable units. For (\ref{eq:30}), the total net revenue of any renewable unit i which includes credits from energy and reserve along with power deviation payment, is non-negative, i.e., $\Gamma^{w*}_{i}\geq 0$.
\end{theorem}\par
\vspace{-0.25em}Please refer to Appendix C for the detailed proof. Finally, from the system operator's perspective, the property of revenue adequacy is established in an expected manner. The merchandise surplus ($\mathrm{MS}$) is calculated as the aggregate payments from consumers minus the sum of expected payments made to generators. The expected payment to generator $i$, denoted as $\Gamma'_i$, is given by the sum of payments for energy $\mathcal{E}_i$, payments for reserves $\mathcal{R}_i$, payments for power deviations $\chi_i$, and expected payments for re-dispatch $\sum_k\epsilon_k\mathcal{D}_{k,i}$. Mathematically, $\Gamma'_i$ can be represented as:
\vspace{-0.5em}\begin{align}\label{eq:32}
 \Gamma'_i = \mathcal{E}_i + \mathcal{R}_i + \chi_i + \sum_{k\in\mathcal{K}} \epsilon_k \mathcal{D}_{k,i}.
\end{align}
\vspace{-1em}\begin{theorem}[Revenue Adequacy]
Under assumption (1)-(4) and the assumption that the optimal clearing quantities and prices are solved from model~(\ref{eq:4}),  merchandise surplus under the proposed settlement is non-negative and is equal to the sum of congestion rent, denoted as,
\begin{align}\label{eq:33}
    \Gamma^d_i-\Gamma^{g'}_i-\Gamma^{w'}_i=\sum_{t \in T} \sum_{k\in \mathcal{K}} f_{k,t}^{\top} \mu_{k, t}^{*}+\sum_{t \in T} f_{t}^{\top} \mu_{t}^{*} \geq 0.
\end{align}
\end{theorem}\par
We give the mathematical proof in Appendix D. For multi-area, how system operators jointly collect congestion rent is elaborated in \cite{b27}. We only discuss the net revenue of the entire system is always non-negative and equal to jointly-covered rent under the optimal dispatch in this paper. 

\vspace{-1em}\section{Numerical Tests}
In this section, the proposed market properties are validated by clearing results from the simple yet illustrative 2-bus system for a single-period simulation and a large system embedded with 3-area and 118-bus for a multi-period simulation. All tests are solved by MATLAB 2022a with the GUROBI 10.0.1 as the solver. 
\vspace{-1em}\subsection{Case 1: a 2-bus example}
\begin{figure}[!ht]
  \centering
  \includegraphics[width=8.7cm,height=2cm]{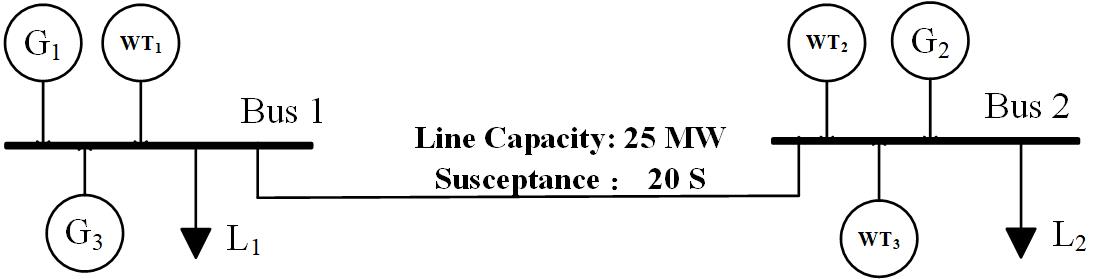}
  \captionsetup{singlelinecheck=off, justification=raggedright}
 \caption{{Network of 2-bus system.}}
  \label{tab:1}
\end{figure}

We use a simple yet illustrative 2-bus system, as depicted in Figure \ref{tab:1}, to validate the aforementioned market properties.
Bus 1 is equipped with a load L1 of 80MW, a wind generator WT1 of 75MW, and two conventional generators: G1 and G3. G1 offers the cheapest energy and reserve options, while G3 offers the most expensive ones. Bus 2 has two wind generators, WT2 and WT3, with capacities of 10MW and 15MW respectively, a conventional generator G2, and a load L2 of 80MW. A transmission line connects bus 1 and bus 2 with a capacity of 25MW. The parameter settings for G1, G2, and G3 are detailed in Table \uppercase\expandafter{\romannumeral1}. 
\begin{table}[!ht]
\caption{Generator Parameters for the 2-Bus System}
\textit{
\begin{center}
    \begin{tabular}{ccccc}
  \hline\hline
    & \rm{Capacity limit (MW)} & \multicolumn{3}{c}{\rm{Bid-in price (\$/MWh)}} \\
    & $G$/$R$   &$c_\mathrm{g}$ &$c_\mathrm{u}/c_\mathrm{d}$ &$\overline{C}/\underline{C}$\\
 \hline
   \rm{G1} & \rm{150 /20} & \rm{2} & \rm{1}& \rm{0.5}  \\
   \rm{G2} & \rm{200 /40} & \rm{4} & \rm{2}& \rm{1}   \\ 
   \rm{G3} & \rm{120 /10} & \rm{6} & \rm{3}& \rm{1.5}   \\ 
   \hline\hline
\end{tabular}
\end{center}}
\end{table}
\vspace{-1em}
We consider a potential G1 outage and renewable and load power deviations in non-base scenarios, as detailed in Table \uppercase\expandafter{\romannumeral2}. The probability of base-case realization was $1-\sum{\epsilon_k}=0.35$ and the base-case RPS is set as $\nu_{1,0}=\nu_{2,0}=0.5$ for each bus.\par
\begin{table}[!ht]
\caption{Non-base Scenarios Settings for the 2-Bus System}
\textit{
\begin{center}
    \begin{tabular}{cccl}
   \hline \hline
    Scenario& Deviation (MW) & Probability& RPS (Bus1,Bus2) \\
   \hline
   S1 & ${L}_{1}$$\downarrow$ 10,${WT}_{1}$$\downarrow$ 15 & 0.15 &(0.5000, 0.6050) \\
   S2 & ${L}_{1}$$\downarrow$ 20,${WT}_{1}$$\downarrow$ 15 & 0.15 &(0.5500, 0.6050) \\
   S3 & ${L}_{1}$$\uparrow$ 10,${WT}_{1}$$\uparrow$ 20 & 0.1 &(0.6050,0.6050) \\
  S4 & ${L}_{1}$$\uparrow$ 20,${WT}_{1}$$\uparrow$ 10 & 0.15&(0.6050,0.6050)  \\
   S5 & ${WT}_{1}$$\downarrow$ 75 & 0.05&(0,0.3125) \\
   S6 & ${L}_{1}$$\uparrow$ 40, ${G}_{1}$ outage & 0.05 &(0.6250,0.3125)\\
   \hline\hline
\end{tabular}
\end{center}}
\end{table}
\par
\vspace{-1em}First, we solve the optimal dispatch model. Clearing results for the base case are listed in Table \uppercase\expandafter{\romannumeral3}, including each generator's energy and reserve clearing quantities and prices.  
\begin{table}[!ht]
\caption{Cleared Energy and Reserve Quantities and Prices for the 2-Bus System}
\begin{center}
\begin{tabular}{ccccccc}
\hline\hline  Generator  & $g$ & $r_{U}$ & $r_{D}$ & $\pi$ & $\pi^{U}$ & $\pi^{D}$ \\
\hline  G1  &15 & 20 & 6.4 & 6 & 2.675 & 1 \\
  G2  &40 & 40 & 0 & 4 & 4.625 & 1.275 \\
  G3  &10 & 10 & 0 & 6 & 6.575 & 1.55 \\
WT1  &70 & 5 & 70 & 6.875 & 6.875 & 0 \\
 WT2  &10 & 0 & 10 & 4.875 & 4.875 & 0 \\
WT3  & 15 & 0 & 15 & 4.875 & 4.875 & 0 \\
\hline\hline
\end{tabular}
\end{center}
\end{table}
\begin{table}[!ht]
\centering
\caption{Validating Theorem 2 for G1 and WT1}
\resizebox{9.5cm}{!}{%
\begin{tabular}{ccccccccccc}
\hline\hline 
Scen & Gen &$\pi_{k}$& $\pi_{k}^{\mathrm{u}}$ & $\pi_{k}^{\mathrm{d}}$ & $\delta x_k$ & $\mathcal{R}_k$ & $\chi_{k}$ & $\epsilon_k\mathcal{D}_k$ & $\pi_{k}\cdot\delta x_k$& $\Pi_k^U+\Pi_k^D$ \\
\multirow[t]{2}{*}{ S1 } & G1 & 0.075 & 0 & 0 & 0 & 0 & 0 & 0 & 0 & 0 \\
 & WT1 & 0.075 & 0.075 & 0 & -10 & 0.375 & -1.125 & 0 & -0.75 & -0.75 \\
\multirow[t]{2}{*}{ S2 } & G1 & -0.875 & 0 & 0.95 & -6.4 & 6.08 & -0.48 & 0 & 5.6 & 5.6 \\
 & WT1 & 0 & 0 & 0 & -13.6 & 0 & 0 & 0 & 0 & 0 \\
\multirow[t]{2}{*}{ S3 } & G1 & 0 & 0 & 0.05 & -6.4 & 0.32 & -0.32 & 0 & 0 & 0 \\
 & WT1 & 0 & 0 & 0 & 16.4 & 0 & 0 & 0 & 0 & 0 \\
 \multirow[t]{2}{*}{ S4 } & G1 & 0.075 & 0 & 0 & 5 & 0 & 0.375 & 0 & 0.375 & 0.375 \\
& WT1 & 0.075 & 0.075 & 0 & 15 & 0.375 & 0.75 & 0 & 1.125 & 1.125 \\
 \multirow[t]{2}{*}{ S5 } & G1 & 2.7 & 2.675 & 0 & 20 & 53.5 & 0.5 & 0 & 54 & 54 \\
& WT1 & 2.7 & 2.7 & 0 & -70 & 13.5 & -202.5 & 0 & -189 & -189 \\
\multirow[t]{2}{*}{ S6 } & G1 & 4.025 & 0 & 0 & -15 & 0 & -0.375 & -60 & -60.375 & -60.375 \\
 & WT1 & 4.025 & 4.025 & 0 & 5 & 20.125 & 0 & 0 & 20.125 & 20.125 \\
\hline\hline
\end{tabular}
}
\end{table}
\par
In the $5^{th}$ column, the energy prices for renewable and conventional units at the same bus are non-uniform because in scenario 2, RPS price $\nu_{1,2}=\nu_{2,2}=0.875 \$$. In scenario 2, if WT1 increases by 1 MW, G1 will curtail 1 MW through re-dispatch. Then in the base case, G1 would provide an extra 1 MW downward reserve at $\pi^{d,g}_2(1)^*=0.95 \$$, and curtailing 1MW in scenario 2 with probability 0.15, resulting in a $0.95-0.15*0.5=0.875 \$$ increase in the total expected system cost. In the last two columns, WT2 and WT3 located at bus 2 have uniform reserve prices. 
\par Next, we validate Theorem 2. Take G1 and WT1 at bus 1 as an example, we record their clearing results in each non-base scenario in Table \uppercase\expandafter{\romannumeral4}. Columns $3^{rd}-5^{th}$ show fractional energy and reserve prices in scenario $k$ for G1 and WT1. The comparison of reserve prices also validates Corollary 1. The power re-dispatch and the product of re-dispatch volume and energy price for each scenario are recorded in the $6^{th}$ and $10^{th}$ columns, respectively. Meanwhile, $7^{th}-9^{th}$ columns record the reserve payment, power deviation payment and expected re-dispatch payment for G1 and WT1. The sum of these payments, denoted as $ \Pi_k^U + \Pi_k^D $, is listed in the last column. The energy prices for G1 and WT1 are non-uniform due to the binding RPS in S2. The equivalence of G1 and WT1 at the same bus is confirmed by the equality between the last two columns for the remaining non-base scenarios. Similar observations can be made to other generators located at the same bus, and we omit the validation here.  
\begin{table}[!ht]
    \centering
    \caption{Settlement in the 2-Bus System [\$].}
    \resizebox{9cm}{!}{%
 \begin{tabular}{ccccccccccc}
\hline\hline  $\Gamma^{g'}_1$ & $\Gamma^{g'}_2$ & $\Gamma^{g'}_3$ & $\Gamma^w_1$ & $\Gamma^w_2$  & $\Gamma^w_3$ & $\Gamma^d_1$  & $\Gamma^d_2$  & $\mathrm{CR}$\\
\hline  89.6 & 349 & 127.25 & 312.75 & 48.75 & 73.125 &  688.125 & 362.35 & 50 \\
\hline 79.75 & 323 & 120.75 & 307.25 & 40 & 60 & 660.75 & 320 & 50 \\
\hline\hline
\end{tabular}}
\end{table}\par
Then, we record the expected payments for each market participant in the first row of Table \uppercase\expandafter{\romannumeral5}, with the revenue for generators in the first six columns, payments for L1 and L2 in  $7^{th}-8^{th}$ columns, and the total congestion rent in the last column. The congestion rent collected by ISO equals the merchandise surplus, which validates revenue adequacy. The settlement outcomes for wind farms supplying only energy are recorded in the second row of Table \uppercase\expandafter{\romannumeral5}. In this case, the costly conventional generator dispatches more reserve, leading to a total expected cost of \$394.75, which is higher than the total expected cost \$391.60 when renewable supply reserve. Additionally, when examining the $4^{th}$ to $6^{th}$ columns, it's evident that the profits received by renewable units are lower when solely providing energy. Additionally, with the realization of any non-base scenario, the profits of G1, G2 and G3 are 53.24 \$, 189 \$, 67.25 \$, the profits of WT1, WT2 and WT3 are 312.75 \$, 48.75 \$, 73.125 \$, which validates cost recovery for participants. 
\vspace{-1em}\subsection{Case 2: a modified 3-area IEEE 118-bus system}
 In this subsection, we perform sensitivity analyses to assess the impact of renewable penetration and power deviation level on the profit of market participants. We evaluate the effectiveness of renewable reserve provision on the IEEE 118-bus test systems proposed in \cite{b28}, and we further modified it to integrate renewable-based generation.
\begin{figure}[t]
  \centering
  \includegraphics[width=7 cm,height=4.5cm]{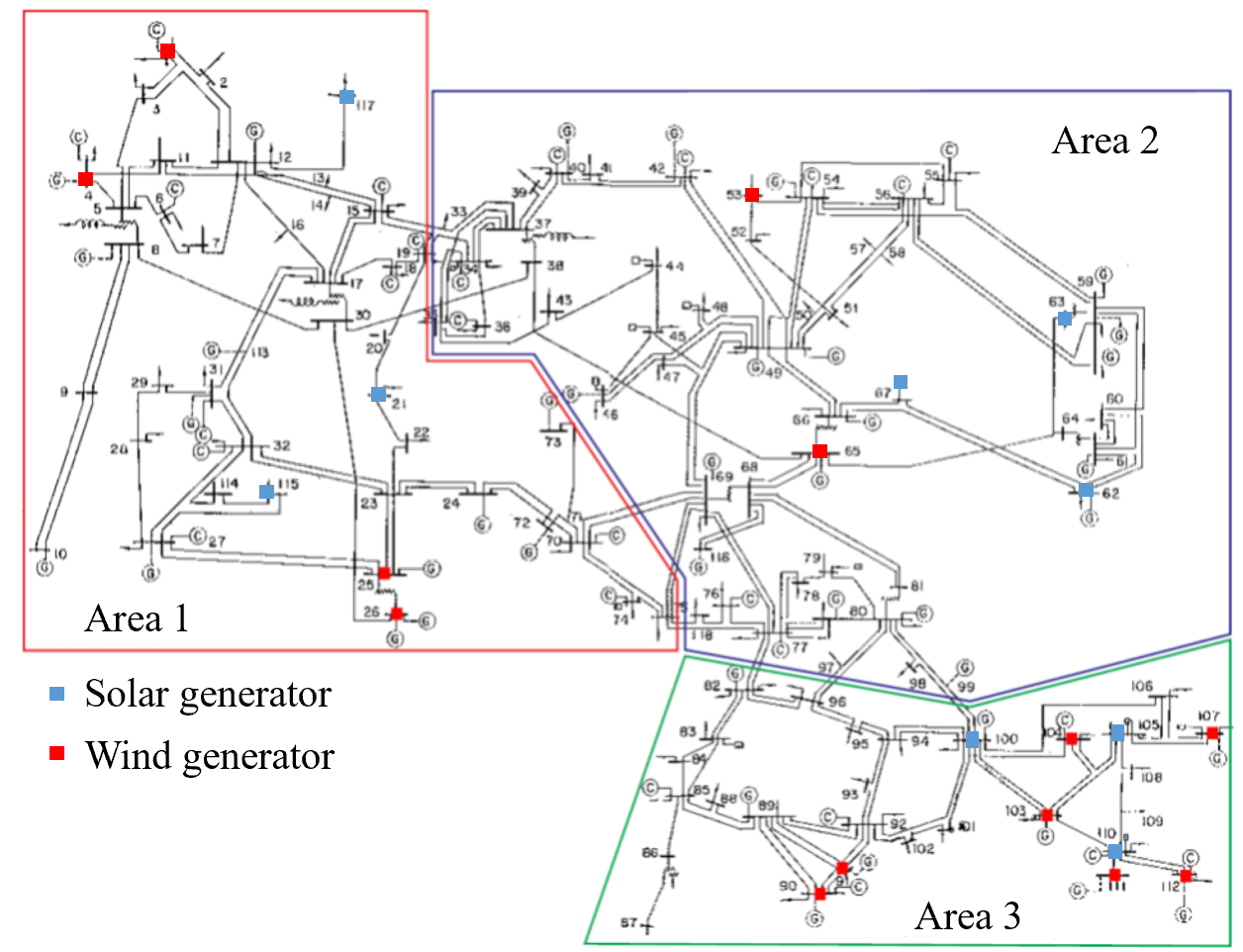}
  \captionsetup{singlelinecheck=off, justification=raggedright}
 \caption{IEEE 118-bus 3-area test system in~\cite{b28}.}
 \label{tab:2}
\end{figure}
As shown in Fig.\ref{tab:2}, the test system contains 91 buses with loads, 54 conventional units, 13 wind generators, and 9 solar generators. The 118-bus system is divided into three regions, with the area partitioning being identical to that in~\cite{b28}. The line parameters remain unchanged except that we adjust the transmission capacities to 100 MW for congestion cases and 1000 MW for non-congestion cases. Conventional generators have energy bids ranging from 20\$ to 74\$, with increments of 1\$. The bid-in price for each generator's re-dispatch, upward, and downward reserves is set at 0.1, 0.2, and 0.2 of its energy bid-in price, respectively. The capacities of conventional generators are set as 0.5 of the original upper bounds specified in MATPOWER \cite{b29}. Both upward and downward reserve capacities are set at 0.3 of the maximum generation capacities, and the ramping rate is fixed at 50 MW per hour.
\begin{figure}[t]
  \centering
\includegraphics[width=9cm,height=2.6cm]{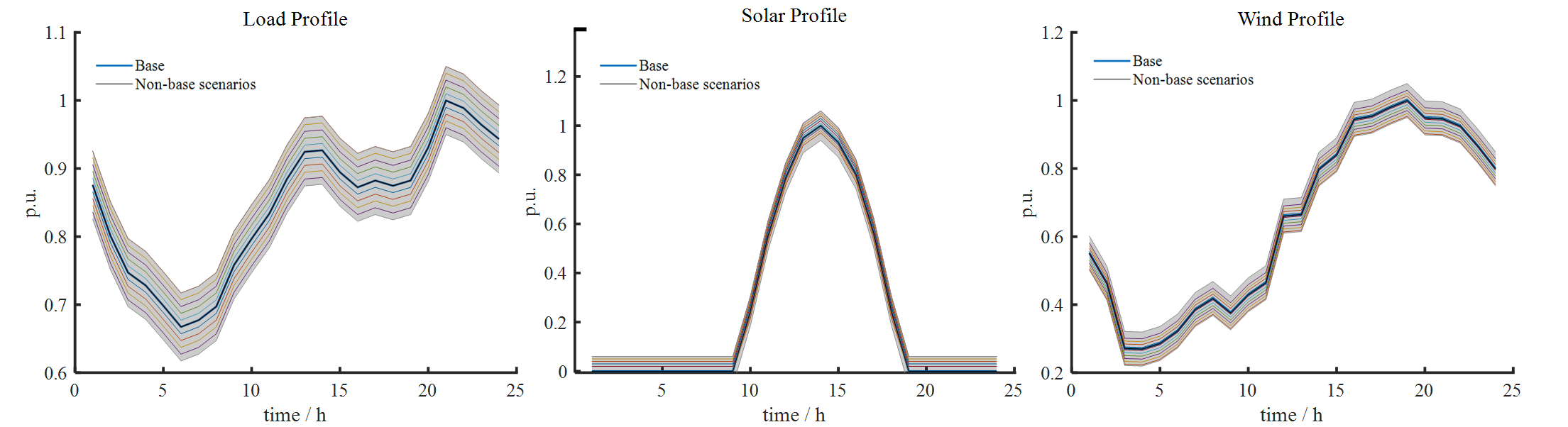}
\captionsetup{singlelinecheck=off, justification=raggedright}
 \caption{Probabilistic forecast value at 5\% deviation levels.}
  \label{tab:3}
\end{figure} 
Detailed hourly time-series load data and simulated wind and solar data are generated from the HRP-38 system~\cite{b30}. The database includes one year of hourly nominal values for wind, solar, and load profiles. Assumed that the basic probability of each scenario is 1/365. Scenario reduction is carried out based on the probability distance, resulting in 10 scenarios and their corresponding probabilities, as shown in Table \uppercase\expandafter{\romannumeral6}. 
\begin{table}[!ht]
    \centering
    \caption{Non-base Scenarios for the 118-Bus Case.}
\resizebox{!}{1.8cm}{\begin{tabular}{ccc}
\hline\hline NO. & Net load deviations $\Delta$ and Outages & Probability \\
\hline 1 &$\Delta$=-0.007, no outage &0.121 \\
2 &$\Delta$=0.002, no outage &0.153\\
3 & $\Delta$=-0.009, no outage&0.051\\
4 &$\Delta$=-0.002, no outage &0.155\\
5 &$\Delta$=0.012, no outage &0.071 \\
6 &$\Delta$=-0.011, G23 at area 2 outage  &0.034 \\
7 &$\Delta$=0.023, no outage &0.073\\
8 &$\Delta$=-0.030, no outage &0.153 \\
9 &$\Delta$=0.007, no outage &0.144 \\
10 &$\Delta$=0.038, G1 at area 1 outage &0.045 \\
\hline\hline
\end{tabular}}
\end{table}\par
The base-case renewable power and load profiles are then derived by calculating the weighted average of these ten scenarios. Subsequently, at each time $t$, the power deviation between the base case and the non-base case follows a nominal distribution within a standard deviation of $k$ portion of the mean value using normalization $ X_{\text {new }}=k\mu \cdot \frac{(X-\mu)}{\sigma}$, where $ X_{\text {new }}$ and $X$ denote new and original value of the data point. $\mu$ denotes mean of the data and $\sigma$ denotes standard deviation of the data. $k$ represents the weight factor and is considered 0.05 in the simulation. Fig.\ref{tab:3} shows scenarios from a scaled profile within 5\% standard deviation of the mean value\footnote{The normalized standard deviation averaged over for all the demand and renewable generators.}. We present the simulation horizon as a day with 24-time slots of 1 hour each.\par
First, we analyze the impact of different energy mix shares on market suppliers' profits. The solar and wind energy penetration in the system is calculated as the ratio of the total base-case renewable energy to the total base-case load over a day. 100\% penetration means that base case renewable generation throughout the day equals the base case load power throughout the day. Two levels of renewable penetration were considered, 50\% and 100\%. A consistent capacity of 20 MW is assigned to the base and non-base nominal profiles of each wind unit, while a capacity of 40 MW is assigned to each solar unit, regardless of penetration level. This setting results in the system being supplied with 4298.8 MWh of daily renewable energy in the base case. The distinct capacity is then applied to the base and non-base per unit load demand for different penetration levels. However, the capacity remains identical for the same penetration level. We set the RPS ratio as at least 0.4 for the base case and 0.2 for all non-base scenarios for each area. 
\begin{figure}[t]
\subfigure[]{\label{tab:4a}\includegraphics[width=4.7cm,height=3.5cm]{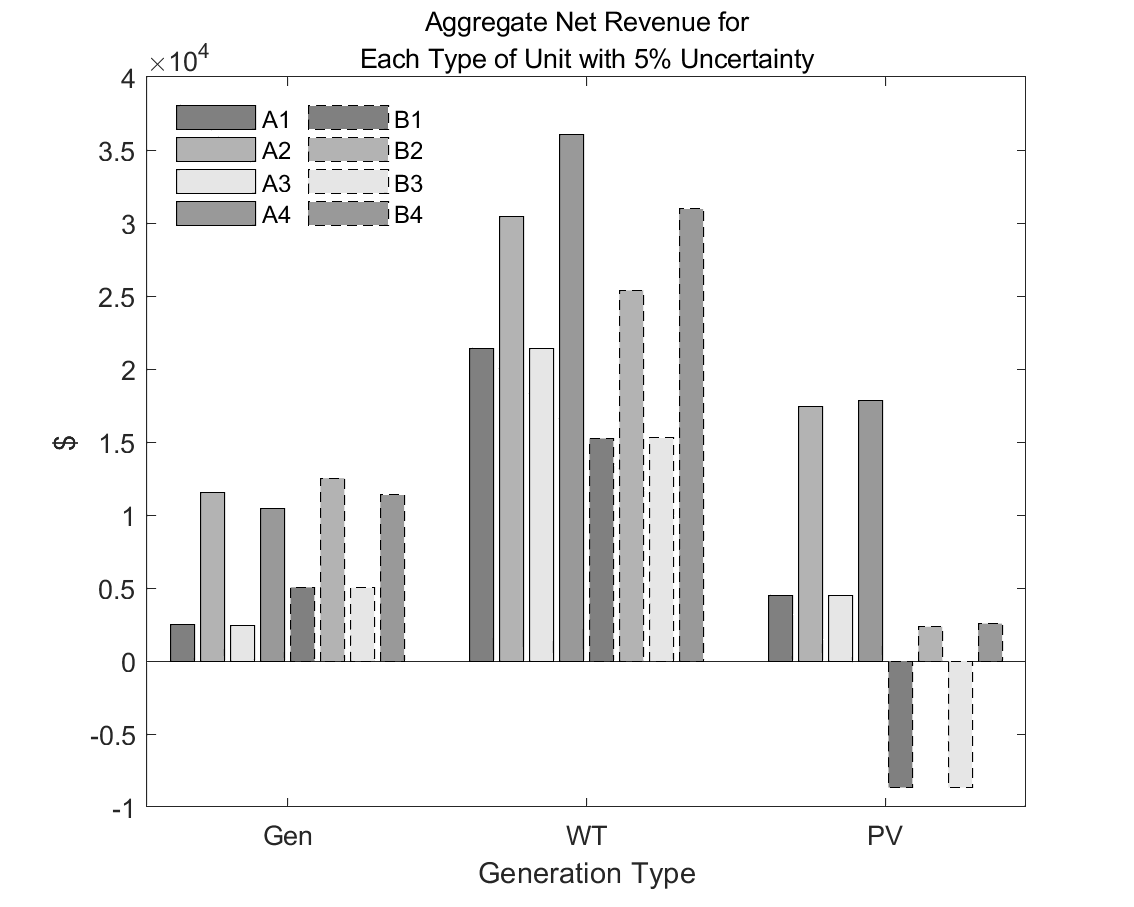}}\hspace{-4pt}
\subfigure[]{\label{tab:4b}\includegraphics[width=4.2cm,height=3.5cm]{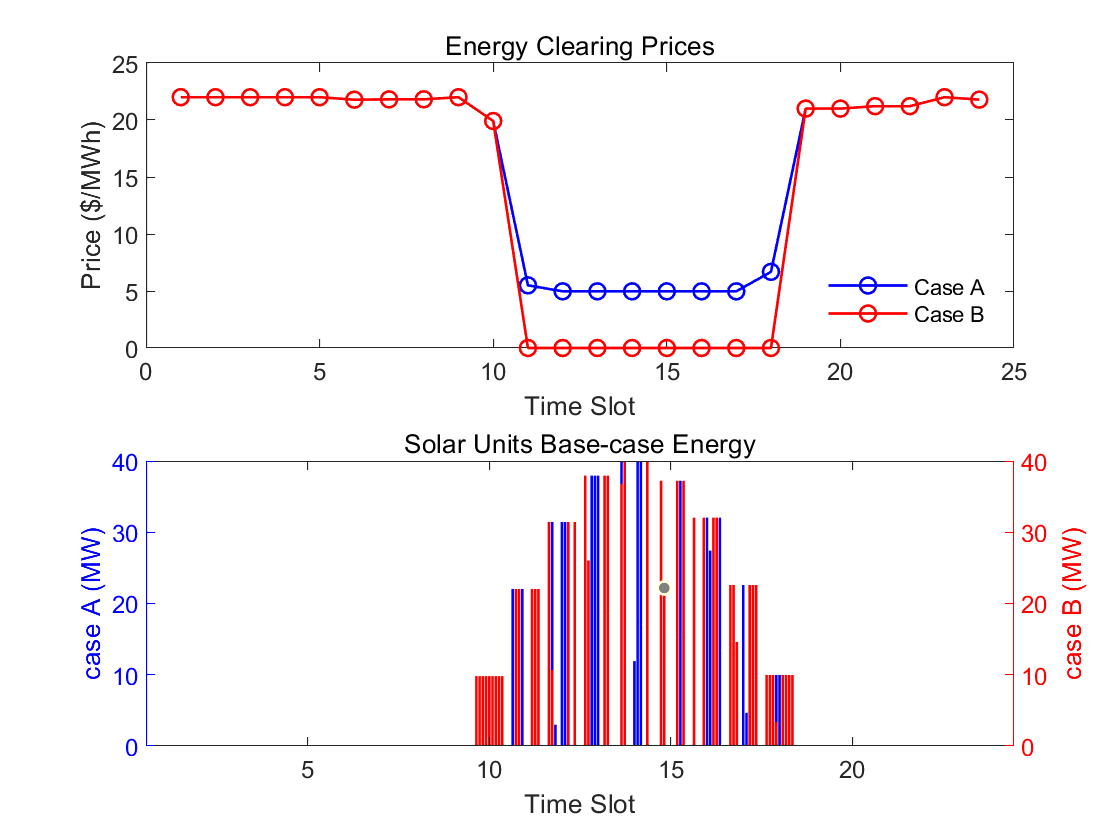}}\hspace{-4pt}
\captionsetup{singlelinecheck=off, justification=raggedright}
\caption{Impact of different renewable penetration levels: (a) Aggregate expected profit of producers; (b) Hourly clearing prices and base-case solar output for Case \textbf{A1} and Case \textbf{B1}.}
\end{figure}
We evaluate the effectiveness of renewable reserve supply by comparing two cases: Case \textbf{A} represents our proposed optimal dispatch model, and Case \textbf{B} considers a system relying solely on conventional generators for reserve supply. In Case \textbf{B}, we modify the optimal dispatch model~(\ref{eq:4}) by introducing infinitely large reserve bid-in costs for RESs to enforce RESs generating energy as much as possible. In each group of tests, we set different line capacities and different renewable penetration ratios, with four cases set as follows, \\
\textbf{\textbf{1}:} 
Renewable penetration is 100\% without line congestion.\\
\textbf{\textbf{2}:} 
Renewable penetration is 50\% without line congestion.\\
\textbf{\textbf{3}:} 
Renewable penetration is 100\% with line congestion.\\
\textbf{\textbf{4}:} 
Renewable penetration is 50\% with line congestion. \par
For a succinct illustration, Figure \ref{tab:4a} presents the aggregated profits of units for each energy type. These profits encompass the sum of individual net revenues from each thermal, wind, and solar unit. Each unit exhibits a non-negative profit in test group A, thereby confirming cost recovery. In each group of tests, thermal units make more profit in Case \textbf{B} than in Case \textbf{A} due to higher inflexible renewable penetration. This results in more binding operating constraints and increased reserve procurement. However, the ability of RESs to provide reserves results in higher profits for wind and solar units, compared to when they only supply energy. We conducted the detailed analysis based on Figure \ref{tab:4b}. Figure \ref{tab:4b} shows the hourly energy prices and the base case output of solar units for Cases \textbf{A1} and \textbf{B1}. Note that the energy price in Case \textbf{B1} consistently drops to zero in the middle of the day. In contrast, due to the priority dispatch of renewable upward reserves, there is a higher level of renewable integration in case \textbf{A1} compared to case \textbf{B1}. Note that renewable upward reserve price $\pi_{k, t}^{\mathrm{u}, \mathrm{w}*}$ is no less than the reserve price of the thermal unit at the same bus, $\pi_{k, t}^{\mathrm{u}, \mathrm{g}*}$. The renewable upward price includes the opportunity cost for not providing reserve and is equal to the energy clearing price, $\pi_{k, t}^{\mathrm{w}*} = \pi_{k, t}^{\mathrm{u}, \mathrm{w}*}$. The energy clearing price remains above zero at noon in Case \textbf{A}. We also record the total system cost in Table \uppercase\expandafter{\romannumeral7}. It is observed that the total system cost in Case \textbf{A} is lower than that in the corresponding scenario in Case \textbf{B}. This indicates that the provision of reserve capacity by RES can reduce flexibility requirements and overall costs, thus providing a more cost-effective method of integrating RES.
\begin{figure}[t]
  \centering
\includegraphics[width=6.5cm,height=3.5cm]{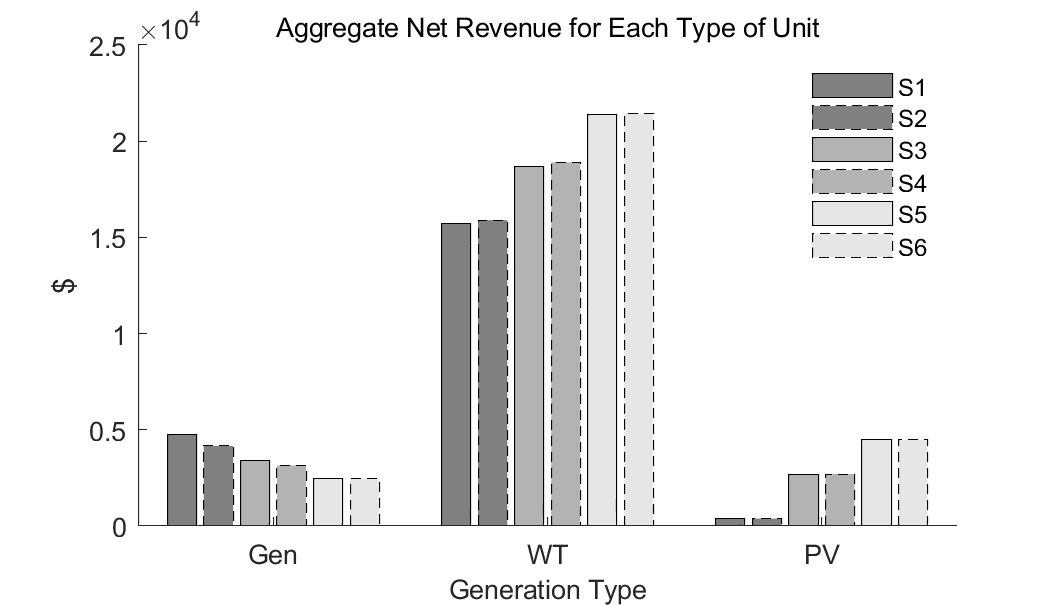}
\captionsetup{singlelinecheck=off, justification=raggedright}
\caption{Aggregate expected profit of producers at different renewable deviation levels.}
  \label{tab:5}
\end{figure}
\begin{table}[t]
    \centering
     \caption{Expected Total Cost in the 118-Bus System [\$].}
     \resizebox{9cm}{!}{
 \begin{tabular}{ccccccccc}
\hline\hline  Case&A1 & A2 & A3 & A4 & B1& B2& B3& B4\\
\hline Cost&40402 & 119243 & 40412 &  125262 & 53195 & 127825 & 53205 & 133845\\
\hline\hline
\end{tabular}}
\end{table}
We next verify the impacts of different uncertainty levels of RES on the market participants' profit. The uncertainty level of load and renewable production can be measured by the standard deviation of the profile normalized by the base-case power curve. The renewable penetration level is fixed at 100\%. The RES standard deviation increases from 5\% to 15\% of the expectation with a 5\% increment. The load standard deviation is equal to 5\% for all cases. In Fig.\ref{tab:5}, S1 and S2, S3 and S4, S5 and S6 represent the total expected system cost when the renewable power deviation level is 15\%, 10\%, and 5\%, respectively, in non-congested and congested conditions. As the uncertainty level increases, conventional unit profits increase due to increased reserve procurement to protect against renewable and load uncertainties. Conversely, there is a decrease in revenue for renewable units, but none of the individual renewable unit profits are negative.
\vspace{-0.5em}\section{Conclusions}
We consider the challenge of limited operational flexibility faced by system operators due to a high share of renewable energy with uncertain output. We investigate the procurement, pricing and settlement in a joint energy-reserve market with renewable portfolio standards and reserve provisions from renewable units. We have developed a 4-part settlement, applicable to both renewable and thermal units, which includes energy, reserve, power deviation and re-dispatch. Our study shows that thermal and renewable generators can be considered equivalent when their respective uncertainties and flexibilities are taken into account. Furthermore, our analysis proposes a cost-effective strategy for integrating renewable generation. We find that allocating capacity for reserve, rather than generating as much energy as possible, can be more economically advantageous and relieve system flexibility requirements. We have also mathematically proven that cost recovery and revenue adequacy are assured under the proposed settlement. A possible limitation of this work is that the proposed clearing model has a high computation burden with large samples. We leave the consideration of extending and solving our optimal dispatch model in a distributed manner for future work.

\ifCLASSOPTIONcaptionsoff
  \newpage
\fi

\appendices
\vspace{-1em}\section{Proof of Theorem 1}
According to the first-order optimal condition with respect to upward power re-dispatch $\delta \mathrm{w}_{k, t}^{+}(i)$ of renewable unit $i$, we have:
\begin{align}\label{eq:34}
\frac{\partial \mathcal{L}}{\partial \delta \mathrm{w}_{k, t}^{+}(i)^{*}}=\overline{\tau_{k, t}}(i)-\underline{\tau_{k, t}}(i)-\pi_{k,t}^{w}(i)= 0.
\end{align}\par
Note that multipliers are associated with inequality constraints that are non-negative. Constraints (\ref{eq:16}) and {(\ref{eq:34})} suggest that the scenario $k's$ fractional contribution of upward reserve price $\pi_{k,t}^{\mathrm{u}, \mathrm{w}}=\overline{\tau_{k, t}}$ and downward reserve price $\pi_{k,t}^{\mathrm{d}, \mathrm{w}}=\overline{\zeta_{k, t}}$ cannot be positive at the same time $t$. i.e.,   
\begin{enumerate}
    \item Net load power deviation is positive in scenario $k$. $\pi_{t}^{\mathrm{d}, \mathrm{w}}=0, \pi_{k,t}^{w}=\pi_{t}^{\mathrm{u}, \mathrm{w}}\geq 0$. 
    \item System is over-generation and power deviation is partly offset by conventional unit in scenario $k$. $\pi_{t}^{\mathrm{u}, \mathrm{w}}=0, \pi_{k,t}^{w}=-\pi_{t}^{\mathrm{d}, \mathrm{w}}\leq 0$.    
    \item Power deviation is totally offset by RES in scenario $k$. $\pi_{k,t}^{w}=-\pi_{t}^{\mathrm{d}, \mathrm{w}}=\pi_{t}^{\mathrm{u}, \mathrm{w}}=0$.
\end{enumerate}
Apparently, for any renewable unit $i$ and renewable unit $j$ located at the same bus, they have uniform energy prices, denoted as $\pi^w_t(i)=\pi^w_t(j)$. Consequently, we can prove the locational-uniform properties for renewable energy reserve prices. The uniform pricing property does not hold if non-zero reserve bid-in costs are applied to renewable resources.
\section{Proof of Theorem 2}
 For renewable unit $i$, by multiplying both sides of (\ref{eq:16}) with $\delta \mathrm{w}_{k, t}^{-}(i)^{*}$ and both sides of (\ref{eq:34}) with $\delta \mathrm{w}_{k, t}^{+}(i)^{*}$, and considering the complementary slackness of (\ref{eq:41}) and (\ref{eq:4}m), we have: 
 \begin{align}
&\hspace{-0.8em}\pi_{k, t}^{\mathrm{u}, \mathrm{w}}(i)^{*}\left(r_{i, t}^{\mathrm{u}, \mathrm{w}}\right)^{*} =\pi_{k,t}^{w}(i)^{*}\delta \mathrm{w}_{k, t}^{+}(i)^{*}-\pi_{k, t}^{\mathrm{u}, \mathrm{w}}(i)^{*}\phi_{k,i,t}^{\mathrm{w},+}, \\
&\hspace{-0.8em}\pi_{k, t}^{\mathrm{d}, \mathrm{w}}(i)^{*}\left(r_{i, t}^{\mathrm{d}, \mathrm{w}}\right)^{*}=\pi_{k, t}^{w}(i)^{*}\phi_{k,i,t}^{\mathrm{w},-}-\pi_{k, t}^{w}(i)^{*}\delta \mathrm{w}_{k, t}^{-}(i)^{*}. 
\end{align}\par
Moreover, since the expected re-dispatch payment for renewable unit $\mathcal{D}^w$ is zero, the left-hand side of (\ref{eq:26}) can be written as:
\begin{align}
    \Pi_{k,t}^{U}(i)+\Pi_{k,t}^{D}(i)&=\pi_{k, t}^{\mathrm{u}, \mathrm{w}}(i)^{*}\left(r_{i, t}^{\mathrm{u}, \mathrm{w}}\right)^{*}+\pi_{k, t}^{\mathrm{u}, \mathrm{w}}(i)^{*}\left(\phi_{k,i,t}^{\mathrm{w},+}\right)\nonumber\\
    &+\pi_{k, t}^{\mathrm{d}, \mathrm{w}}(i)^{*}\left(r_{i, t}^{\mathrm{d}, \mathrm{w}}\right)^{*}-\pi_{k, t}^{w}(i)^{*}\left(\phi_{k,i,t}^{\mathrm{w},-}\right)\nonumber\\
    &=\pi_{k, t}^{\mathrm{u}, \mathrm{w}}(i)^{*}\left(\delta \mathrm{w}_{k, t}^{+}(i)^{*}-\delta \mathrm{w}_{k, t}^{-}(i)^{*}\right),\label{eq:37}
\end{align}
which equals to the right-hand side of (\ref{eq:26}).
Similarly, we can derive (\ref{eq:37}) for renewable unit $j$ located at the same bus. Therefore, we can prove theorem 2 for renewable units.\par
For conventional units $i$ and $j$ at the same bus, suppose unit $i$ is not shut down and unit $j$ is outage in scenario $k$. For unit $i$, by setting the partial Lagrangian $\mathcal{L}$ with respect to upward power deviation $\delta \mathrm{g}_{k, t}^{+}(i)^{*}$, we show that
\begin{align}
\hspace{-1em}\frac{\partial \mathcal{L}}{\partial\delta \mathrm{g}_{k, t}^{+}(i)^{*}}=\epsilon_{k} \bar{C}(i)+\overline{\eta_{k, t}}(i)^{*}-\underline{\eta_{k, t}}(i)^{*}-\pi_{k, t}^{\mathrm{g}}(i)^{*}= 0.\label{eq:38}
\end{align}\par
Multiplying both sides of (\ref{eq:9}) by $\delta \mathrm{g}_{k, t}^{-}(i)^{*}$ and both sides of (\ref{eq:38}) by $\delta \mathrm{g}_{k, t}^{+}(i)^{*}$, and taking into account the complementary slackness condition of constraints (\ref{eq:4}j) to (\ref{eq:4}k), we can derive (\ref{eq:39}) and (\ref{eq:40}) as follows:
\begin{align}
&\hspace{-1.2em}\pi_{k, t}^{\mathrm{u}, \mathrm{g}}(i)^{*}\left(r_{i,t}^{\mathrm{u}, \mathrm{g}}\right)^{*}+\epsilon_{k}\bar{C}(i)\delta \mathrm{g}_{k,t}^{+}(i)^{*}=\pi_{k, t}^{g}(i)^{*}\delta \mathrm{g}_{k,t}^{+}(i)^{*},\label{eq:39} \\
&\hspace{-1.2em}\pi_{k, t}^{\mathrm{d}, \mathrm{g}}(i)^{*}\left(r_{i, t}^{\mathrm{d}, \mathrm{g}}\right)^{*}-\epsilon_{k} \underline{C}(i)\delta \mathrm{g}_{k,t}^{-}(i)^{*}= -\pi_{k, t}^{g}(i)^{*}\delta \mathrm{g}_{k,t}^{-}(i)^{*}.\label{eq:40}
\end{align}\par
By substituting (\ref{eq:39})-(\ref{eq:40}) into the left-hand side of (\ref{eq:26}) and considering that the power deviation payment $\chi_k^{g,-}(i)=0$, we have:
\begin{align}
    \Pi_{k,t}^{U}(i)+\Pi_{k,t}^{D}(i)&=\pi_{k, t}^{\mathrm{u}, \mathrm{g}}(i)^{*}\left(r_{i, t}^{\mathrm{u}, \mathrm{g}}\right)^{*}+\pi_{k, t}^{\mathrm{d}, \mathrm{g}}(i)^{*}\left(r_{i, t}^{\mathrm{d}, \mathrm{g}}\right)^{*}\nonumber\\
    &+\epsilon_{k}\bar{C}(i)\delta \mathrm{g}_{k,t}^{+}(i)^{*}-\epsilon_{k}\underline{C}(i)\delta \mathrm{g}_{k,t}^{-}(i)^{*}\nonumber\\
    &=\pi_{k, t}^{\mathrm{g}}(i)^{*}\left(\delta \mathrm{g}_{k, t}^{+}(i)^{*}-\delta \mathrm{g}_{k, t}^{-}(i)^{*}\right).
\end{align}\par
For unit $j$, both upward and downward power deviations equal to 0, $\delta \mathrm{g}_{k, t}^{+}(j)^{*}=0$ and $\delta \mathrm{g}_{k, t}^{-}(j)^{*}=g_{j,t}^{*}$, consequently, the combination payment is reformulated as:
\begin{align}\label{eq:41}
    &\Pi_{k,t}^{U}(j)+\Pi_{k,t}^{D}(j)=-\epsilon_{k}\underline{C}(j)\delta \mathrm{g}_{k,t}^{-}(j)^{*}+\chi^g_j\nonumber\\&=\pi_{k, t}^{\mathrm{g}}(j)^{*}\left(-\delta \mathrm{g}_{k, t}^{-}(j)^{*}\right)=\pi_{k, t}^{\mathrm{g}}(j)^{*}\left(\delta \mathrm{g}_{k, t}^{+}(j)^{*}-\delta \mathrm{g}_{k, t}^{-}(j)^{*}\right).
\end{align}\par
Along with $\pi_{k, t}^{\mathrm{u}, \mathrm{g}}(j)=\pi_{k, t}^{\mathrm{u}, \mathrm{g}}(i)$, we can prove theorem 2 for conventional units regardless of whether there are outages. When RPS constraints are inactive, $nu=0$, we may easily extend this proof to any renewable unit $i$ and conventional generator $j$ on the same bus,  $\pi_{k, t}^{\mathrm{w}}(i)=\pi_{k, t}^{\mathrm{g}}(j)$.\par
Furthermore, note that green premium $\nu$ and re-dispatch bid-in cost $\bar{C},\underline{C}$ are non-negative. Consider any conventional unit $i$ and renewable generator $j$ at the same bus, (\ref{eq:9}), (\ref{eq:16}), (\ref{eq:34}) and (\ref{eq:38}) indicate that
\begin{align}
&\hspace{-1em}\pi_{k, t}^{\mathrm{u}, \mathrm{g}}(i)^{*}=\pi_{k, t}^{\mathrm{u}, \mathrm{w}}(j)^{*}-\epsilon_{k} \overline{C}(i)-\nu_{k}^{*}(m_j),\forall k \in \mathcal{K}, i \notin \Omega_{K} \\
&\hspace{-1em}\pi_{k, t}^{\mathrm{d}, \mathrm{g}}(i)^{*}= \pi_{k, t}^{\mathrm{d}, \mathrm{w}}(j)^{*}+\epsilon_{k} \underline{C}(i)+\nu_{k}^{*}(m_j), \forall k \in \mathcal{K}, i \notin \Omega_{K}.
\end{align}\par
When conventional unit $i$ is shut down in scenario $k$,  the system is more likely to face insufficient upward reserve, as indicated by $i \in \Omega_{K}$. Based on the discussion in Appendix A, renewable upward and downward reserve prices cannot be positive at the same time, i.e., $\pi_{k, t}^{\mathrm{u}, \mathrm{w}}(j)^{*}\geq\pi_{k, t}^{\mathrm{u}, \mathrm{g}}(i)^{*}=0$ and $\pi_{k, t}^{\mathrm{d}, \mathrm{w}}(i)^{*}=\pi_{k, t}^{\mathrm{d}, \mathrm{g}}(i)^{*}=0$. Taking into account all non-base scenarios, it is observed that renewable unit $j$ has a lower downward reserve price and higher upward reserve price than conventional unit $i$, as delineated in Corollary 1. 
\section{Proof of Theorem 4}
Within this proof, we will focus on a particular renewable generator, say renewable unit $i$. For brevity, we drop the subscript $i$ and superscript $*$ of all primal and dual variables associated with renewable generator $i$. Setting the partial derivatives of $\mathcal{L}$ with respect to optimal energy $w_{t}$, upward reserve $r^{\mathrm{u}, \mathrm{w}}_{t}$ and downward reserve $r^{\mathrm{d}, \mathrm{w}}_{t}$ equal to zero, we obtain the conditions for optimality: 
\vspace{-0.5em}\begin{align}
&\frac{\partial \mathcal{L}}{\partial w_{t}} =0 \implies-\pi_{t}^{w}+\overline{\iota ^{U}_{t}}-\overline{\iota ^{D}_{t}}=0,\\
&\frac{\partial \mathcal{L}}{\partial r^{\mathrm{u}, \mathrm{w}}_{t}} =0 \implies-\pi_{t}^{\mathrm{u}, \mathrm{w}}+ \overline{\iota ^{U}_{t}}-\underline{\iota ^{U}_{t}}=0,\label{eq:46}\\
&\frac{\partial \mathcal{L}}{\partial r^{\mathrm{d}, \mathrm{w}}_{t}} =0 \implies -\pi_{t}^{\mathrm{d}, \mathrm{w}}+ \overline{\iota ^{D}_{t}}-\underline{\iota ^{D}_{t}}=0.
\end{align}\par
Combined with the complementary slackness condition of (\ref{eq:4}c), we can easily reformulate the combination of energy and reserve credits as: 
\begin{align}\label{eq:48}
   \pi^w_{t}  w_{t}+\pi_{t}^{\mathrm{u}, \mathrm{w}}  r^{\mathrm{u}, \mathrm{w}}_{t}+\pi_{t}^{\mathrm{d}, \mathrm{w}}  r^{\mathrm{d}, \mathrm{w}}_{t}=\bar{l}_t^U \bar{W}_t.
\end{align}\par
Consider $\phi_{k, t}^{\mathrm{w},-}$ is a scenario-dependent parameter. Thus, it cannot be extracted from a summation operation. Also, note that $0\leq\phi_{k, t}^{\mathrm{w},-}\leq\overline{W}_t, \forall k \in \mathcal{K}$ and $\overline{W}_t $ remains constant across scenarios. We therefore reduce the power deviation payment $\chi_{i}^{w}$ defined in (\ref{eq:22}) as the right-hand side of (\ref{eq:49}):
\begin{align}\label{eq:49}
    \chi_{i}^{w}\geq\sum_{k \in \mathcal{K}}\left(\pi_{k, t}^{\mathrm{u}, \mathrm{w}} \phi_{k, t}^{\mathrm{w},+}-\pi_{k, t}^{\mathrm{w}} \bar{W}_t\right).
\end{align}\par
Combined with (\ref{eq:48}) and (\ref{eq:49}), the objective function of profit-maximization model in (\ref{eq:30}) can be reformulated as:
\vspace{-1em}\begin{align}\label{eq:50}
 \Gamma^w & \geq\sum_{t \in \mathcal{T}}\left(\bar{l}_t^U \bar{W}_t+\sum_{k \in \mathcal{K}}\left(\pi_{k, t}^{\mathrm{u}, \mathrm{w}} \phi_{k, t}^{\mathrm{w},+}-\pi_{k, t}^{\mathrm{w}} \bar{W}_t\right)\right)\nonumber \\
& =  \sum_{t \in \mathcal{T}}\left(\left(\underline{l}_t^U+\pi_t^{\mathrm{u}, \mathrm{w}}\right) \bar{W}_t+\sum_{k \in \mathcal{K}}\left(\pi_{k, t}^{\mathrm{u}, \mathrm{w}} \phi_{k, t}^{\mathrm{w},+}-\pi_{k, t}^{\mathrm{w}} \bar{W}_t\right)\right) \nonumber\\
& = \sum_{t \in \mathcal{T}}\left(\underline{l}_t^U \bar{W}_t+\sum_{k \in \mathcal{K}}\left(\pi_{k, t}^{\mathrm{u}, \mathrm{w}} \phi_{k, t}^{\mathrm{w},+}+\left(\pi_{k, t}^{\mathrm{u}, \mathrm{w}}-\pi_{k, t}^{\mathrm{w}}\right) \bar{W}_t\right)\right)\nonumber\\
&  = \sum_{t \in \mathcal{T}}\left(\underline{l}_t^U\bar{W}_t+\sum_{k \in \mathcal{K}}\left(\overline \tau_{k, t} \phi_{k, t}^{\mathrm{w},+}+\underline \tau_{k, t} \bar{W}_t\right)\right),
\end{align}
where the first equality comes from the optimality condition (\ref{eq:46}), the second equality is derived by distributive property. The last equality comes from (\ref{eq:34}). The right-hand side of (\ref{eq:50}) sums up products of non-negative parameters and dual variables. Thus, the left-hand side of (\ref{eq:50}) is non-negative. Given that energy payment $\mathcal{E}^w_i$, reserve payment $\mathcal{R}^w_i$, and power deviation payment $\chi^w_i$ are settled in the ex-ante stage, the profit with every scenario realization is equal to $\Gamma^w$ and always non-negative. Consequently, we can prove the cost recovery for renewable units.\par
Next, we provide the mathematical proof for the conventional units. For brevity, we drop the superscript $*$ of all primal and dual variables associated with conventional generator $i$. Setting the partial derivatives of $\mathcal{L}$ with respect to optimal energy $g_{i,t}$, upward reserve $r^{\mathrm{u}, \mathrm{g}}_{i,t}$ and downward reserve $r^{\mathrm{d}, \mathrm{g}}_{i,t}$ equal to zero, the optimal solutions of optimal dispatch model~(\ref{eq:4}) also satisfies the KKTs conditions of profit-maximization problem~(\ref{eq:29}):
\begin{align}
\frac{\partial \mathcal{L}}{\partial g_{i,t}}=&c_{\mathrm{g}}(i)-\pi_{t}^{g}(i)+\sum_{k \in \mathcal{K},i \in \Omega_{K}}\pi_{k,t}^{g}(i)\nonumber\\
&+\bar{\imath_{t}}(i)-\underline{\imath_{t}}(i)+\Delta\gamma_{t}^{U}(i)-\Delta\gamma_{t}^{D}(i)=0, \\
\frac{\partial \mathcal{L}}{\partial r_{i,t}^{\mathrm{u}, \mathrm{g}}}=& c_{\mathrm{u}}(i)-\pi_{t}^{\mathrm{u}, \mathrm{g}}(i)+\bar{\imath_{t}}(i)+\overline{\ell_{t}^{U}}(i)-\underline{\ell_{t}^{U}}(i)+\gamma_{t}^{U}(i)=0, \end{align}
\begin{align}
 \frac{\partial \mathcal{L}}{\partial r_{i,t}^{\mathrm{d}, \mathrm{g}}}=&c_{\mathrm{d}}(i)-\pi_{t}^{\mathrm{d}, \mathrm{g}}(i)+\underline{\imath_{t}}(i)+\overline{\ell_{t}^{D}}(i)-\underline{\ell_{t}^{D}}(i)+\gamma_{t}^{D}(i)=0,   
\end{align}
where $\Delta\gamma_{t}^{U}=\gamma_{t}^{U}-\gamma_{t-1}^{U}$ and $\Delta\gamma_{t}^{D}=\gamma_{t}^{D}-\gamma_{t-1}^{D}$. Combined with the complementary slackness conditions of constraints (\ref{eq:4}d)-(\ref{eq:4}f), for any conventional unit $i$, the net revenue can be reformulated into:
\begin{align}\label{eq:54}
\Gamma^g_i-\Xi^g_i=&\sum_{t\in \mathcal{T}}\Big(\big(\pi_{0,t}^{g}(i)+\sum_{k \in \mathcal{K},i \notin \Omega_{K}}\pi_{k,t}^{g}(i)-c_{\mathrm{g}(i)}\big) g_{i,t}\nonumber\\
&+\left(\pi_{t}^{\mathrm{u}, \mathrm{g}}(i)-c_{\mathrm{u}}(i)\Big) r_{i,t}^{\mathrm{u}, \mathrm{g}}+\left(\pi_{t}^{\mathrm{d}, \mathrm{g}}(i)-c_{\mathrm{d}}(i)\right) r_{i,t}^{\mathrm{d}, \mathrm{g}}\right) \nonumber\\
=& \sum_{t\in \mathcal{T}}\Big(\bar{\imath}_{t}(i)\overline{G_i}+\overline{\ell^{U}_{t}}(i)R_i^{\mathrm{u},\mathrm{g}}+\overline{\ell^{D}_{t}}(i)R_i^{\mathrm{d},\mathrm{g}}\nonumber\\
&+\gamma^{D}_{t}(i)\Delta g_i^{D}+\gamma^{U}_{t}(i)\Delta g_i^{U} \Big),
\end{align}
where the right-hand side of~(\ref{eq:54}) includes the summation of several parts, each one is a product of a non-negative parameter and a dual variable. Therefore the left-hand side of ~(\ref{eq:54}) is necessarily non-negative. Namely, we can prove cost recovery for conventional units.
\vspace{-1em}\section{Proof of Theorem 5}
To prove the revenue adequacy for the system operator, for the base case, we first rewrite the power balance equation and transmission capacity constraints (\ref{eq:4}b) in phase angle-based formation:
\begin{align}
(\lambda^{\prime}_{t},\mu_{t})& : B\theta = g_{t}+w_{t}-d_{t}, F \theta \leq f,
\end{align}
where $\theta$ denotes the vector of the phase angles and $\lambda_{t}^{\prime}$ represents price vector of nodal energy marginal prices. The matrix $B$ denotes the bus admittance matrix and matrix $F$ represents the branch admittance matrix. According to stationary conditions associated with the phase angle-based optimal dispatch model, 
\begin{align}\label{eq:56}
\frac{\partial L}{\partial \theta^{*}} &=B^{\top} \lambda^{\prime *}_{t}+F^{\top} \mu^{*}_{t}=0, 
\end{align}
with phase angle $\theta^{*}$ multiplied to both left-hand and right-hand sides of~(\ref{eq:56}),
\begin{align}
& {\theta^{*}}^{\top} \frac{\partial L}{\partial \theta^{*}} =(B \theta^{*})^{\top} \lambda_{t}^{\prime *}+(F \theta^{*})^{\top} \mu_{t}^{*} \nonumber \\
& =\left(g_{t}^{*}+w_{t}^{*}-d_{t}\right)^{\top} \lambda_{t}^{\prime *}+f^{\top} \mu_{t}^{*}=0. 
\end{align} \par
Combined with the complementary slackness condition of base-case RPS (\ref{eq:4}g), we can exactly proof revenue adequacy in the base case.
\begin{align}\label{eq:58}
    \hspace{-1em}\mathrm{MS_0}=\sum_{t \in \mathcal{T}}\Big({\pi_{0,t}^{d}}^{\top}{d}_{t}-{\pi_{0,t}^{w}}^{\top}w_{t}^{*}-{\pi_{0,t}^{g}}^{\top}g_{t}^{*}\Big)=\sum_{t \in \mathcal{T}}f_{t}^{\top} \mu_{t}^{*}.
\end{align}\par
Similarly, we derive the complementary slackness condition for constraints (\ref{eq:4}h)-(\ref{eq:4}i) in any scenario $k \in \mathcal{K}$ at time $t$
\begin{small}
\begin{align}\label{eq:59}
\left(\begin{array}{l}
{\pi^d_{k,t}}^{\top}(d_{t}+\phi_{k,t}^{\mathrm{d}})-\\
{\pi^w_{k,t}}^{\top}\left({w}_{t}^{*}+(\delta \mathrm{w}_{k, t}^{+})^{*}-(\delta \mathrm{w}_{k, t}^{-})^{*}\right)-\\
{\pi^g_{k,t}}^{\top}\left({g}_{t}^{*}+(\delta \mathrm{g}_{k, t}^{+})^{*}-(\delta \mathrm{g}_{k, t}^{-})^{*}\right)
\end{array}\right)=f_{k,t}^{\top} \mu_{k,t}^{*}.
\end{align}
\end{small}\par
By substituting constraint~(\ref{eq:26}) into the 4-part expected payment (\ref{eq:32}) for conventional and renewable units, combined with the load payment, we can compute the expectation of merchandise surplus as follows:
\begin{align}
\mathrm{MS} = \sum_{t \in T} \sum_{k\in \mathcal{K}} f_{k}^{\top} \mu_{k, t}^{*}+\sum_{t \in T} f_{t}^{\top} \mu_{t}^{*}\geq 0.
\end{align}\par
With the base-case revenue adequacy proved in (\ref{eq:58}) and the revenue adequacy for any non-base scenario in (\ref{eq:59}), we can prove the revenue adequacy proposed in (\ref{eq:33}).

\end{document}